\documentclass[amsmath,amssymb,aps,prapplied, twocolumn,superscriptaddress, longbibliography,floatfix]{revtex4-2}

\RequirePackage{epsfig}
\RequirePackage{graphicx}

\RequirePackage{xcolor}
\RequirePackage{siunitx}
\RequirePackage{nicefrac}

\RequirePackage{booktabs}
\RequirePackage{makecell}

\RequirePackage{enumerate}
\RequirePackage{enumitem}

\RequirePackage[normalem]{ulem}
\RequirePackage{wrapfig}
\RequirePackage{verbatim}

\usepackage{todonotes} % for comments

\newcommand{\CD}[1]{{\color{pink}#1}}

\newlength{\wdth}

\RequirePackage[colorlinks=true,linkcolor=blue,urlcolor=blue,hyperfootnotes=false,citecolor=black,pdfencoding=auto,psdextra]{hyperref}
\RequirePackage{hypernat} % makes hyperref play nice with natbib
\allowdisplaybreaks[1]
\graphicspath{{../Figures/}}

\RequirePackage[capitalise]{cleveref}

\newcommand{\sinc}{\mathrm{sinc}}

\begin{document}

\title{Magnetic-field dependence of a Josephson traveling-wave parametric amplifier and integration into a high-field setup}

\author{L.~M.~Janssen}
\affiliation{Physics Institute II, University of Cologne, Zülpicher Str. 77, 50937 Köln, Germany}
\author{G. Butseraen}
\affiliation{Université Grenoble Alpes, CNRS, Grenoble INP, Institut Néel, 38000 Grenoble, France}
\author{J.~Krause}
\affiliation{Physics Institute II, University of Cologne, Zülpicher Str. 77, 50937 Köln, Germany}
\author{A. Coissard}
\affiliation{Silent Waves, 25 Avenue des Martyrs, 38000 Grenoble, France}
\author{L.~Planat}
\affiliation{Silent Waves, 25 Avenue des Martyrs, 38000 Grenoble, France}
\author{N. Roch}
\affiliation{Université Grenoble Alpes, CNRS, Grenoble INP, Institut Néel, 38000 Grenoble, France}
\affiliation{Silent Waves, 25 Avenue des Martyrs, 38000 Grenoble, France}
\author{G. Catelani}
\affiliation{JARA Institute for Quantum Information (PGI-11), Forschungszentrum Jülich, 52425 Jülich, Germany}
\affiliation{Quantum Research Center, Technology Innovation Institute,  Abu Dhabi 9639, UAE}
\author{Yoichi~Ando}
\affiliation{Physics Institute II, University of Cologne, Zülpicher Str. 77, 50937 Köln, Germany}
\author{C.~Dickel}
\email[correspondence should be addressed to: \newline]{ando@ph2.uni-koeln.de \newline dickel@ph2.uni-koeln.de}
\affiliation{Physics Institute II, University of Cologne, Zülpicher Str. 77, 50937 Köln, Germany}

\begin{abstract}
We investigate the effect of magnetic field on a photonic-crystal Josephson traveling-wave parametric amplifier (TWPA). 
We show that the observed change in photonic bandgap and plasma frequency of the TWPA can be modeled by considering the suppression of the critical current in the Josephson junctions (JJs) of the TWPA due to the Fraunhofer effect and closing of the superconducting gap. 
Accounting for the JJ geometry is crucial for understanding the field dependence. 
In one in-plane direction, the TWPA bandgap can be shifted by \SI{2}{\giga\hertz} using up to \SI{60}{\milli\tesla} of field, without losing gain or bandwidth, showing that TWPAs without SQUIDs can be field tunable.
In the other in-plane direction, the magnetic field is perpendicular to the larger side of the Josephson junctions, so the Fraunhofer effect has a smaller period.
This larger side of the JJs is modulated to create the bandgap. The field interacts more strongly with the larger junctions, and as a result, the  TWPA bandgap closes and reopens as the field increases, causing the TWPA to become severely compromised already at \SI{2}{\milli\tesla}. 
A slightly higher operating limit of  \SI{5}{\milli\tesla} is found in out-of-plane field, for which the TWPA's response is hysteretic.
These measurements reveal the requirements for magnetic shielding needed to use TWPAs in experiments where high fields at the sample are required; we show that with magnetic shields we can operate the TWPA while applying over \SI{2}{\tesla} to the sample.
\end{abstract}

\maketitle

\section{Introduction}
\label{sec:introduction}

Superconducting parametric amplifiers have become a key tool in quantum technology because they enable low-noise readout of weak microwave signals~\cite{esposito2021}. 
Nowadays, thanks to phase matching techniques, Traveling-Wave Parametric Amplifiers (TWPAs) can combine high gain, bandwidth, and saturation power~\cite{Macklin15, Planat2020, HoEom2012, Goldstein2020, Malnou2021, Shu2021, Ranadive2022}, positioning themselves as ideal tools for scalable low-noise amplification.
However, since most TWPAs depend on superconductivity, they face inherent problems in the presence of strong magnetic fields. 
This is particularly problematic because many experiments involving magnetic fields at the sample could benefit from or already rely on the low-noise amplification provided by TWPAs.
These include superconducting qubit experiments \cite{Luthi18, Schneider19, Pita-Vidal20, Kringhoj21, Krause22}, experiments with spin ensembles \cite{Bienfait2016, Wang2023}, spin qubits \cite{Schaal2020, deJong2021, Vine2023} or topological qubits~\cite{aghaee2024}, as well as hybrid setups including mechanical~\cite{Kounalakis2020, Bera21} and magnonic degrees of freedom~\cite{Yutaka2015, Kounalakis2022}, and the search for axions~\cite{Bartram2023, DiVora2023}.
While magnetic-field-resilient parametric amplifiers made from high-kinetic-inductance superconductors have been demonstrated recently~\cite{xu2023, Frasca2023, Splitthoff2023}, \mbox{TWPAs} resilient to magnetic fields are, to our knowledge, yet to be demonstrated. 
There are now several recent experiments with magnetic fields at the sample that have utilized TWPAs, e.g. \cite{Uilhoorn2021, Wang2023, Bartram2023, DiVora2023, Elhomsy2023, aghaee2024}, emphasizing the need for a comprehensive study of TWPAs in magnetic field.

\begin{figure}
  \centering
  \includegraphics[width=\columnwidth]{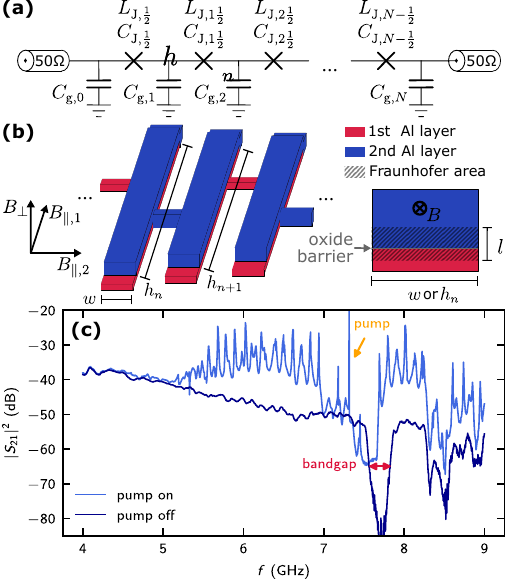}
%  \end{center}
    \caption{
    \textbf{(a)} Schematic of the TWPA electrical circuit. The series of Josephson junctions (JJs) forms a non-linear meta-material. The capacitances to ground provide the $\SI{50}{\ohm}$ impedance matching. 
    The Josephson inductance and ground capacitance are periodically modulated for dispersion engineering.
    \textbf{(b)} Schematic of a part of the JJ array of the TWPA (not to scale). Two layers of aluminum with aluminum oxide junction barriers form a series of junctions. The width  $w = \SI{0.7}{\micro\meter}$ of these junctions is the same for all, while the height $h$ is modulated by $\pm\SI{5}{\percent}$ to engineer a photonic bandgap (on average $h = \SI{16}{\micro\meter}$). 
    The field directions are named based on their orientation relative to the junction array.
    On the right, a cross section of the JJ area is shown to illustrate the Fraunhofer area in case magnetic field penetrates the JJ.
    \textbf{(c)} TWPA transmission with the pump tone turned off and on. The gap feature of the TWPA is visible in both signals.
    The dips beyond \SI{8}{\giga\hertz} are not a feature of the TWPA but due to the circulators in the setup being out of specification.
    }
  \label{fig:fig1}
\end{figure}

Here we present measurements and a theoretical description 
of the effect of magnetic field on
a photonic-crystal Josephson-junction (JJ) TWPA~\cite{Planat2020}.
We estimate the maximum field that can be applied to the TWPA without compromising performance, giving an indication of the required magnetic shielding for a given stray field from the device under test.
In the process, we show that in-plane field along an appropriate direction can be a viable tuning knob to change the operating frequency of a TWPA without severely compromising performance.
The tuning is a consequence of the suppression of the critical current of the individual JJs, which is due to the Fraunhofer effect and the suppression of the superconducting gap (see e.g. Ref.~\cite{Krause22}); the former is strongly dependent on JJ geometry.
Due to the distinct values of the critical field and Fraunhofer period, we can tell apart the two contributions to the tuning of the JJs, which was difficult in previous experiments~\cite{Schneider19, Krause22}. 
We also observe deviations from the commonly-used Ginzburg-Landau (GL) formula for the gap suppression, deviations that are consistent with the predictions of Abrikosov-Gorkov (AG) theory~\cite{Abrikosov1961, Skalski1964} for pair-breaking due to magnetic field.
In addition to the magnetic field dependence, we show that the TWPA can be operated in our setup with fields above \SI{2}{\tesla} at the device under test with several layers of magnetic shielding around the TWPA.
We also establish new understanding of the operation of cQED and hybrid experiments in magnetic fields, e.g. if one wants to couple a JJ array to another system in the presence of small magnetic fields.

Two TWPAs are used in this work, each of the commercially available Argo model, made by Silent Waves.
The data in \cref{sec:introduction}, \cref{sec:twpa_vs_Bpar1} and \cref{sec:twpa_vs_Bpar2_Bperp} is measured on one such device, which we call TWPA A, while the data in \cref{sec:shielded} was measured on a second TWPA, TWPA B.
A detailed description and a summary of device parameters can be found in \cref{sec:device_description}.
The circuit diagram and JJ geometry are shown in \cref{fig:fig1} \textbf{(a)}-\textbf{(b)}.
The amplification is based on four-wave mixing, where two pump photons are converted to a signal and an idler photon via a nonlinear material -- in this case a JJ transmission line. 
The JJs are made with the bridge-free double-shadow evaporation technique~\cite{Lecocq2011}.
The JJ width $w$ is designed to be the same for all JJs.
The JJ geometry is varied by modulating the JJ height $h$, to engineer the dispersion relation, which is essential for increasing the gain and bandwidth of photonic-crystal TWPAs.
The JJ geometry is essential for modeling the experiment because one of the two ways magnetic field can influence the Josephson inductance $L_\mathrm{J}$ is via the Fraunhofer effect. 
If a uniform current distribution is assumed, this effect modulates the inductance with a $1/|\sinc(BA/\Phi_\mathrm{0})|$ dependence, where $A$ is the area of the flux-penetrated region of the JJ as shown in \cref{fig:fig1} \textbf{(b)} and $\Phi_0$ is the magnetic flux quantum.
Our data shows that not just the approximately~\SI{2}{\nano\meter} oxide barrier itself, but also some portion of the superconducting film on both sides is penetrated by magnetic field, as expected because of the finite penetration depth for magnetic fields in a superconductor.

The transmission $S_\mathrm{21}$ of the TWPA without a pump and with an optimized pump are shown in \cref{fig:fig1} \textbf{(c)}.
{The measurement setup is described in \cref{sec:measurement_setup}.
The no-pump data shows the photonic bandgap at frequency $f_\mathrm{g}$ which results from the dispersion engineering.   
When a continuous microwave pump tone at the right frequency and power is applied, the transmission around the pump is enhanced due to the four-wave mixing, showing gain over a certain bandwidth on both sides of the bandgap feature.
We will focus on the gain below $f_\mathrm{g}$, as in our setup $f_\mathrm{g}$ is close to the upper edge of the working frequency range of the circulators that were used.

\section{TWPA operation under $B_\mathrm{\parallel,1}$ field}
\label{sec:twpa_vs_Bpar1}

\begin{figure}
  \centering
  \includegraphics[width=\columnwidth]{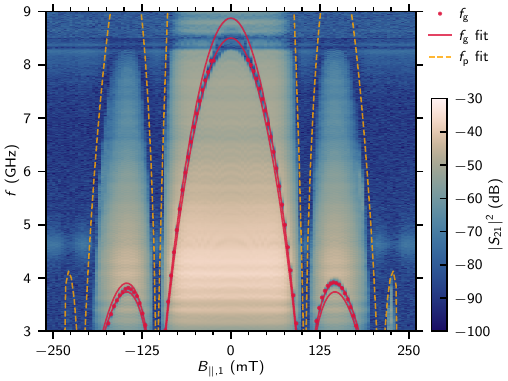}
%  \end{center}{}
  \caption{
    TWPA transmission spectrum without pump tone as a function $B_\mathrm{\parallel,1}$.
    The gap frequency $f_\mathrm{g}$ is extracted wherever it is visible (red dots). 
    The upper and lower edge of the gap are calculated from our model, and shown as solid red lines. 
    The best fit is obtained by taking the average of these two and fitting this to the measured $f_\mathrm{g}$.
    The plasma frequency $f_\mathrm{p}$, based on the same model, is plotted as an orange dashed line.
}
  \label{fig:fig2}
\end{figure}

To understand the magnetic field dependence of the TWPA we start with the $B_\mathrm{\parallel,1}$ direction [as defined in \cref{fig:fig1} (\textbf{b})]. 
In \cref{fig:fig2}, the transmission through the TWPA for a range of field values $B_\mathrm{\parallel,1}$ is plotted.
The two distinctive features of the TWPA transmission are the frequency of the bandgap $f_\mathrm{g}$ and the plasma frequency $f_\mathrm{p}$, where the transmission is cut off.
A Fraunhofer pattern is clearly visible for both.
For an unknown reason, $f_\mathrm{g}$ is slightly different for positive and negative magnetic field, and the second Fraunhofer lobe appears more strongly for the positive field direction.
Hysteresis was considered as the possible reason, but we see the same result for sweeps in both directions.
Therefore we are only able to fit the data accurately for one side.

Starting from the circuit model described in Ref.~\cite{Planat2020}, we can calculate the $f_\mathrm{g}$ and  $f_\mathrm{p}$ for each field value.
The TWPA is a transmission line made of JJs and capacitances to ground as illustrated in \cref{fig:fig1} \textbf{(a)}.
The JJs can be modeled with a Josephson inductance $L_\mathrm{J,n+1/2}$ and capacitance $C_\mathrm{J,n+1/2}$, both of which depend on the junction area.
The Josephson inductance is inversely proportional to the critical current which in turn is proportional to the area. 
The Josephson capacitance is simply proportional to the area.
Thus the modulation of $h$ leads to a modulation of the inductance $L_\mathrm{J,n+1/2}$, capacitance $C_\mathrm{J,n+1/2}$, and the capacitance to ground $C_\mathrm{g,n}$ where the index $n$ denotes the position in the JJ array.
The capacitance to ground $C_\mathrm{g,n}$ of each island is achieved by covering the entire JJ array with dielectric and depositing a thick copper ground plane on top.
Since the area of an island is almost entirely determined by the area of the junctions, we assume the same modulation for $C_\mathrm{g,n}$ as for $C_\mathrm{J,n+1/2}$.

The combination of the modulation of $L_\mathrm{J,n+1/2}$, $C_\mathrm{J,n+1/2}$, and $C_\mathrm{g,n}$ causes a photonic bandgap which is used for phase matching the TWPA.
From this circuit model, one can calculate an effective impedance $Z$ for the TWPA, as well as $f_\mathrm{g}$ and  $f_\mathrm{p}$. 
To calculate the magnetic field dependence of the TWPA transmission and the parameters mentioned above, we need to model the magnetic field dependence of $L_\mathrm{J,n+1/2}$.

We consider that the magnetic field affects the JJs by suppressing the superconducting gap $\Delta$ according to AG theory and through the Fraunhofer effect.
We assume that the capacitances $C_\mathrm{J,n+1/2}$ and $C_\mathrm{g,n}$ are unaffected by the magnetic field.
The Josephson inductances are inversely proportional to the gap: $L_\mathrm{J,n+1/2} \propto \nicefrac{1}{\Delta(B_\mathrm{\parallel,1})}$.
For the Fraunhofer dependence, the geometry matters.
The Fraunhofer effect arises from an interplay of the JJ current and the magnetic field. 
The most straightforward assumption would be a rectangular current distribution in each junction.
However, we find that parametrizing the current distribution to allow for more current density at the edges of the junctions gives a better fit.
It is certainly possible that the oxide barrier is slightly thinner away from the center of the junctions, which could mean that the actual current distribution is indeed closer to the one we use for fitting.
The field $B_\mathrm{\parallel,1}$ is in-plane with respect to the junction array and perpendicular to the narrow side of the junctions of width $w$.
Since $w$ is unchanged throughout the array, the suppression of the critical current is identical for all junctions.
This significantly simplifies the magnetic field modeling: instead of applying a change to each junction individually, we only need to change the average Josephson inductance $\bar{L}_\mathrm{J}$.

The full model can be found in \cref{sec:twpa_model};
the best fit obtained using equations \eqref{eq:plasmaB1}, \eqref{eq:sincreplacement}, and \eqref{eq:bandgap1}, and using as fit parameters the field $B_\mathrm{\Phi,1}$ corresponding to a flux quantum through area $A$, the average Josephson inductance at zero field $\bar{L}_\mathrm{J}$ and the parameter governing the current distribution $\chi$, is shown in \cref{fig:fig2}.
The critical field $B_\mathrm{c}$ is not used as a fit parameter because it can be read-off directly from the data.
As the model also suggests, the magnetic field changes the TWPA impedance, such that in the Fraunhofer minimum $\vert S_{21}\vert$ breaks down because the TWPA becomes reflective.
A TWPA $\vert S_{11}\vert$ measurement can be found in \cref{sec:TWPA_transmission_vs_reflection}.
The features of the measured plasma frequency are reproduced accurately.
The fact that the second Fraunhofer lobe is reproduced indicates that we have a good understanding of the field dependence of the superconducting gap.

\begin{figure}
  \centering
  \includegraphics[width=\columnwidth]{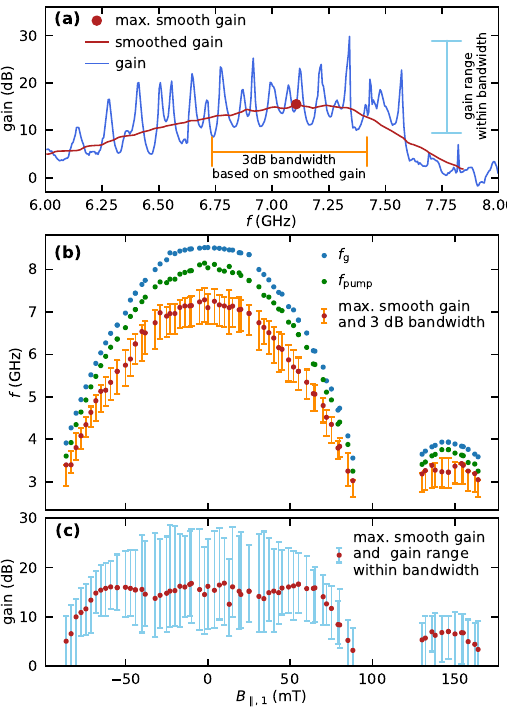}
%  \end{center}
  \caption{
    \textbf{(a)} TWPA gain as a function of frequency. All gain curves are calculated relative to a fixed background estimate, not to the pump-off spectrum. 
    To extract figures of merit that are robust to the gain spikes, we smooth the gain to define the maximum smooth gain and bandwidth.
    \textbf{(b)} frequency of maximum smoothed gain and bandwidth window as a function of $B_\mathrm{\parallel,1}$. 
    The TWPA bandgap feature can be tuned by about $\SI{2}{\giga\hertz}$ with the in-plane field while the optimum gain remains roughly stable, then the gain becomes compromised. 
    \textbf{(c)} maximum smooth gain and gain range within \SI{3}{\decibel}-bandwidth window as a function of $B_\mathrm{\parallel,1}$.
    While the highest gain values are continuously reduced with  $B_\mathrm{\parallel,1}$, the maximum smooth gain remains stable up to about $\vert B_\mathrm{\parallel,1} \vert < \SI{60}{\milli\tesla}$.
    }
  \label{fig:fig3}
\end{figure}

After understanding the effect of the magnetic field based on the pump-off transmission data and circuit model, we now proceed to look at how the amplifier figures of merit, in particular gain and bandwidth, change with $B_\mathrm{\parallel,1}$ for the TWPA pumped with a microwave tone of frequency $f_\mathrm{pump}$ and power $P_\mathrm{pump}$.
An example of the TWPA gain profile is shown in \cref{fig:fig3}~\textbf{(a)}.
We estimate the gain profile not by subtracting the pump-off transmission, but rather by estimating a field-independent background without the bandgap (see \cref{sec:twpa_figs_of_merit} for details on the background subtraction).
The gain profile shows strong fluctuations that depend on $f_\mathrm{pump}$ and $P_\mathrm{pump}$.
In an actual experiment, if one wants to optimize the amplifier gain at a particular frequency, the $f_\mathrm{pump}$ and $P_\mathrm{pump}$ can be optimized to give the best signal-to-noise ratio (SNR) at that particular frequency. 
We have instead decided to smooth the gain profile with a \SI{500}{\mega\hertz} window filter and optimized on the maximum smooth gain [see \cref{fig:fig3} \textbf{(a)}]. The bandwidth of the smoothed gain is also shown as a figure of merit, but not used for the optimization of the pump tone.

We have tuned up the TWPA at different fields by optimizing the maximum smooth gain varying $f_\mathrm{pump}$ and $P_\mathrm{pump}$ as free parameters using the Nelder-Mead method~\cite{Nelder1965}.
\cref{fig:fig3} \textbf{(b)} and \textbf{(c)} show the resulting gain and bandwidth.
More details on the optimization and other figures of merit are reported in \cref{sec:twpa_figs_of_merit}.
We see that the maximum smooth gain remains stable up to $B_\mathrm{\parallel,1}=\SI{60}{\milli\tesla}$, although the maximum gain does decrease.
Additionally, the bandwidth of the gain does not decrease significantly up to this field.
However, the frequency range where amplification occurs is shifted by \SI{2}{\giga\hertz}.
This shows that an in-plane field in this specific direction can be used to effectively tune the amplification range of the TWPA across a wide range.

\section{TWPA operation for other field directions}
\label{sec:twpa_vs_Bpar2_Bperp}

\begin{figure*}
  \centering
      \includegraphics[width=\textwidth]{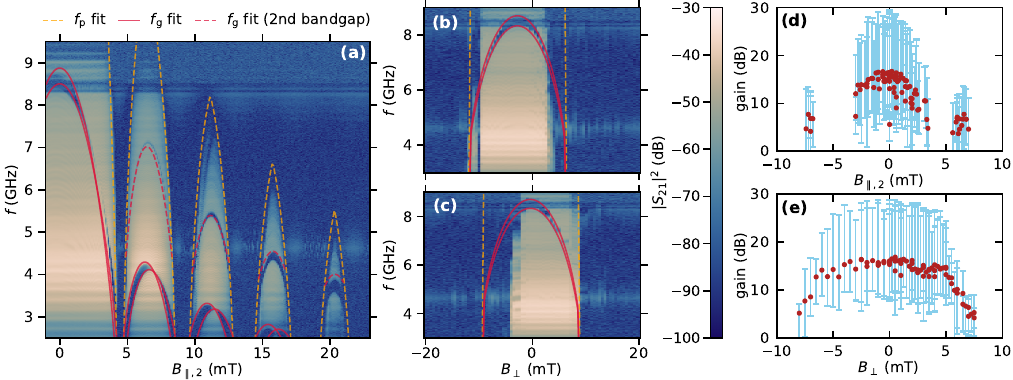}
%  \end{center}
  \caption{
    \textbf{(a)} TWPA transmission spectrum without pump tone as a function $B_\mathrm{\parallel,2}$.
    The edges of the fitted first bandgap $f_\mathrm{g}$ are shown as red lines. The corresponding $f_\mathrm{p}$ is shown as an orange dashed line. 
    The fitted center of the second bandgap is indicated with a dashed red line. 
    The model parameters are the same as in \cref{fig:fig2}, except for the field corresponding to a flux quantum, as the flux now penetrates the longer side of the junction.
    \textbf{(b)} and \textbf{(c)} \CD{no-pump} TWPA transmission spectrum as a function $B_\mathrm{\perp}$ scanning in negative and positive direction respectively.
    Here too, $f_\mathrm{p}$ and $f_\mathrm{g}$ are shown, fitted with a GL-like phenomenological formula (see \cref{sec:Bcrit_par_vs_perp}). 
    The hysteresis is much larger than for the in-plane directions.
    \textbf{(d)} and \textbf{(e)} optimized maximum smooth gain and gain range as a function of $B_\mathrm{\parallel,2}$ and $B_\mathrm{\perp}$ respectively.
  }
  \label{fig:fig4}
\end{figure*}

In this section, we show how the TWPA responds to magnetic fields applied in the other principal field directions $B_\mathrm{\parallel,2}$ and $B_\mathrm\perp$, as indicated in \cref{fig:fig1} \textbf{(b)}, starting with $B_{\parallel,2}$.
The pump-off TWPA transmission as a function of $B_\mathrm{\parallel,2}$ are shown in \cref{fig:fig4} \textbf{(a)}.
There is a Fraunhofer pattern similar to \cref{fig:fig2} but with a much smaller period, indicating a lower $B_{\Phi}$.
Specifically we find $B_{\Phi,2} = \SI{4.55}{\milli\tesla}$, whereas $B_{\Phi,1} = \SI{107.8}{\milli\tesla}$.
This is in-line with the TWPA dimensions, as $\nicefrac{h}{w}=22.7$ which is similar to $\nicefrac{\SI{107.8}{\milli\tesla}}{\SI{4.55}{\milli\tesla}}=23.7$.
This shows that both datasets imply the thickness of the field penetrated area $l$ [see \cref{fig:fig1} \textbf{(b)}] to be the same, namely \SI{28.5}{\nano\meter}.
This value for $l$ is reasonable since it should be compared to roughly half of the total thickness of the two aluminum layers forming the JJs~\cite{Barone}.

There are striking qualitative differences between the dependence on $B_\mathrm{\parallel,2}$ and on $B_\mathrm{\parallel,1}$.
In  \cref{fig:fig4} \textbf{(a)}, we observe a periodic closing and reopening of the bandgap. 
This is due to the interplay between the modulation of $h$, which creates the bandgap, and the Fraunhofer effect.
In the $B_\mathrm{\parallel,1}$ direction, $w$ is nominally identical for all JJs, so we should obtain the same Fraunhofer dependence for all JJs resulting in a common prefactor for all $L_{\mathrm{J},n+1/2}$ (assuming perfectly identical geometries), as mentioned previously. 
But in $B_\mathrm{\parallel,2}$ direction, the different $h_\mathrm{n}$ result in different Fraunhofer contributions: the Fraunhofer critical current modulation has a smaller period for JJs with larger $h$ than for JJs with smaller $h$. 
However, the JJs with large $h$ had lower $L_\mathrm{J}$ to begin with. 
As a result, the difference between the $L_\mathrm{J}$ values for the different junctions decreases initially with increasing field, and at some point, this effectively cancels out the modulation of the capacitances \CD{$C_{\mathrm{J},n+1/2}$} (which remains constant with field) and closes the bandgap.
More details on this modeling, based on the same combination of AG gap suppression and Fraunhofer effect as for $B_\mathrm{\parallel,1}$, can be found in \cref{app:Bpar2}.
We also see higher harmonics of the bandgap in the Fraunhofer side lobes, that are not there for the $B_\mathrm{\parallel,1}$ direction. 
Lastly, there is a line of low transmission with a large upward slope in the second side lobe starting around \SI{10}{\milli\tesla}, but we do not understand this feature.

The $B_\mathrm{\perp}$ field dependence [see \cref{fig:fig4} \textbf{(b)} for sweeping $B_\mathrm{\perp}$ in the negative and \textbf{(c)} for the positive sweep direction] is different from the in-plane directions. 
Here, we initially observe a decrease of $f_\mathrm{g}$, but the transmission then suddenly breaks down.
We see strong hysteresis in the sweep direction: the maximum of $f_\mathrm{g}$ is offset by \SI{2.4}{\milli\tesla}.
This is in strong contrast to the $B_\mathrm{\parallel,1}$ and $B_\mathrm{\parallel,2}$ direction, where we estimate the hysteresis to be less than \SI{1}{\milli\tesla}.
The hysteresis could be due to Abrikosov vortices that an out-of-plane field can create in superconducting films.
The vortices can lead to increased inductance because of the associated currents, but they will also increase dissipation.
We estimate a critical field $B_\mathrm{c,\perp}=\SI{10.3}{\milli\tesla}$ based on the $f_\mathrm{g}$ dependence.
The ratio of the in-plane and out-of-plane critical field is $B_\mathrm{c,\perp}/B_\mathrm{c, \parallel}\simeq0.04$ which is reasonable for thin film aluminum.
A quantitative understanding of vortices in the JJ array is beyond the scope of this work.
More details on the $B_\perp$ direction can be found in  \cref{sec:Bcrit_par_vs_perp}.

We compare the optimized gain as a function of $B_{\parallel,2}$ and $B_{\perp}$ in \cref{fig:fig4} \textbf{(d)} and \textbf{(e)}. 
Notably, the gain collapses for $B_{\parallel,2}$ and $B_\perp$ at \SI{2}{\milli\tesla} and \SI{5}{\milli\tesla}, respectively, while it is stable up to \SI{60}{\milli\tesla} for $B_{\parallel,1}$.
This means that the Fraunhofer dependence for the large JJ direction is imposing the strongest field limit on TWPA performance, but the limits due to vortices and gap suppression in $B_\perp$ are of the same order of magnitude.
The gain and bandwidth for all three different field directions as a function of the estimated $f_\mathrm{g}$ are also 
discussed in \cref{sec:twpa_figs_of_merit}.

\section{Shielded TWPA operation}
\label{sec:shielded}

From the previous measurements,  one can estimate the magnetic shielding required to operate a TWPA in a dilution refrigerator with strong magnetic fields at the device under test.
The cryostat used in this experiment features a \SI{6}{\tesla}/\SI{1}{\tesla}/\SI{1}{\tesla} vector magnet.  
If the TWPA is mounted on the mixing chamber stage, the stray fields at \SI{6}{\tesla} will be on the order of \SI{30}{\milli\tesla}, which is incompatible with TWPA operation. 
The magnetic shielding needs to keep the field at the TWPA below \SI{1}{\milli\tesla}.
For TWPAs with SQUIDs, the shielding requirements are likely more stringent, because the SQUID periods are often chosen to be on the order of \SI{100}{\micro\tesla}.
A strong magnet that is not operated in persistent-current mode can also lead to considerable low-frequency noise~\cite{Krause22} that would likely compromise a SQUID-based TWPA.
To shield the TPWA from magnetic fields, a superconductor is a clear candidate material. 
Aluminum was readily available, and its low critical temperature means it can easily be heated out of the superconducting state and cooled back down again at zero field to regain shielding after being compromised by a field exceeding its critical field. 
However, the critical field of aluminum is not high enough to withstand the large stray field of the vector magnet. 
To mitigate this, outer shields of a high-permeability material can be used. In this experiment, Cryoperm was used, which is a material similar to mu-metal, but designed to perform at cryogenic temperatures.
The shield setup is described in detail in \cref{sec:TWPA_in_fridge_setup}, and shown in \cref{fig:twpa_in_fridge}.

To test the magnetic shields, we measured the transmission through the TWPA and also the optimized gain as a function of the field at the magnet center. 
If the shielding setup were perfect, no change in the transmission or gain would be observed. 
We optimized the gain at zero field and ramped both \SI{1}{\tesla} coils ($B_\mathrm{x}$ and $B_\mathrm{y}$ in the magnet coordinate system) of the vector magnet to the maximum and saw no change in $f_\mathrm{g}$ or in the gain. 
However, the shields are not sufficient to fully shield the stray fields of the \SI{6}{\tesla} magnet $B_\mathrm{z}$, which likely saturates the Cryoperm shield which also leads to the failure of the superconducting aluminum shield. 
As shown in \cref{fig:fig5} \textbf{(a)}, when the magnet is ramped above \SI{2.5}{\tesla}, enough stray field penetrates the shields to change $f_\mathrm{g}$ and quickly suppress the gain [\cref{fig:fig5} \textbf{(a)}]. 
Additionally, we observe ripples in the transmission that become stronger with field.
These ripples are not constant in frequency and we believe them to be a response of the TWPA to low-frequency noise related to the magnet, possibly due to vibrations of the TWPA relative to the magnet.
For the dataset in  \cref{fig:fig5}, we did not optimize the pump settings and ramped back down from \SI{6}{\tesla} to see if the gain would recover, which it did [see \cref{fig:fig5} \textbf{(b)}].

A wide range scan from \SI{6}{\tesla} to \SI{-6}{\tesla} and additional details can be found in \cref{sec:TWPA_in_fridge_setup}.
There we show that we can still get high gain at fields up to \SI{3.5}{\tesla}, but only by running the gain optimization procedure again, since $f_\mathrm{g}$ shifts the pump settings need to be adjusted.
Between \SI{4}{\tesla} and \SI{5}{\tesla}, the TWPA transmission eventually breaks down, showing that the shielding can still be improved. 
However, sweeping the magnet back down, TWPA transmission is restored with a slight shift in $f_\mathrm{g}$. We confirmed that a thermal cycle above \SI{1.5}{\kelvin} can reset the device and the shielding capabilities, showing that the shields are not permanently compromised.
To have more flexibility and be able to measure even with a compromised TWPA, we installed cryogenic switches that allow bypassing the TWPA (see \cref{sec:TWPA_in_fridge_setup}).

\begin{figure}
  \centering
      \includegraphics[width=\columnwidth]{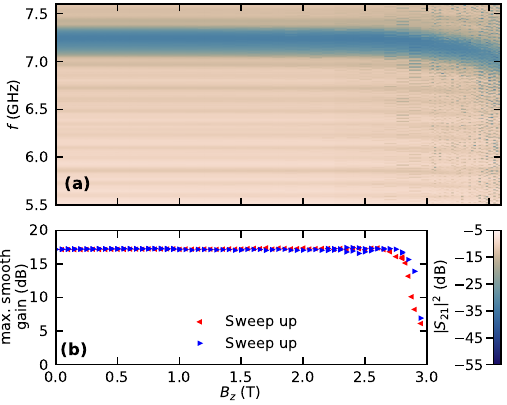}
%  \end{center}
  \caption{
    \textbf{(a)} No-pump transmission of the shielded TWPA as a function of the magnetic field at magnet center.
    Up to about \SI{2.5}{\tesla}, not much changes, but for higher field, the bandgap is tuned down and there is additional noise that manifests at dips in the signal.
    \textbf{(b)} Maximum smooth gain as a function of magnetic field.
    Here the gain was optimized at zero field and settings were not updated at every field. 
    Around \SI{2.5}{\tesla} the gain changes and is strongly reduced by \SI{3}{\tesla}.
    }
  \label{fig:fig5}
\end{figure}

To use the TWPA with higher fields at the sample, one could either improve the shielding or increase the distance between TWPA and magnet. 
In our case, the magnet is below the mixing chamber and we simply mounted the shielded TWPA above at a distance of about \SI{40}{\centi\meter} from the magnet center. 
While one could mount the TWPA higher up and still attach it to the mixing chamber, we investigated the temperature dependence of the TWPA's performance to explore the possibility of mounting it at a higher stage of the dilution refrigerator (see \cref{sec:TWPA_vs_T}).  
We only measured the gain and transmission and found no change in performance up to about \SI{0.3}{\kelvin}. 
While we did not perform detailed measurements of the added noise as a function of temperature, our data would suggest that mounting the TWPA at the cold plate ($T\approx\SI{0.1}{\kelvin}$) is a viable option.

\section{Conclusion}
\label{sec:conlcusion}

In conclusion, we characterized the magnetic field dependence of a Josephson junction TWPA and found that it strongly depends on field direction. 
In-plane magnetic field along the direction perpendicular to the junction array can be an option for tuning the frequency of the amplification band of a JJ TWPA by \SI{2}{\giga\hertz} at similar gain and bandwidth with fields up to \SI{60}{\milli\tesla}.
Magnetic-field tuning of a TWPA with \SI{100}{\micro\tesla} out-of-plane fields has been shown previously in TWPAs with SQUIDs~\cite{Planat2020}.
In-plane field is an interesting alternative as the field required for tuning is about two orders of magnitude larger.
Hence, a similar level of flux noise would lead to less added noise.
Similar ideas for the tuning of JJ-arrays but not in the context of parametric amplifiers have been reported in Ref.~\onlinecite{Kuzmin2023}.
At the same time, a few millitesla of out-of-plane field or in-plane field along the JJ array can severely compromise the TWPA.
We can model the TWPA in-plane field dependence based on the magnetic field dependence of the JJ inductances due to the Fraunhofer effect and the suppression of the superconducting gap.
This suggests that field dependence can be a diagnostic tool for extracting the parameters of the underlying circuit from the model, as the field doesn't affect the capacitances.

Regarding the use of the TWPA in a setup with high magnetic fields at a sample, with magnetic shields and about \SI{40}{\centi\meter} distance between TWPA and magnet center, we show that we can operate the TWPA at more than \SI{2}{\tesla} at the sample.
Between \SI{4}{\tesla} and \SI{5}{\tesla}, stray fields break superconductivity in the TWPA, but after sweeping the field back down and thermal cycling the TWPA recovers.
We also implemented cryo-switches to bypass the TWPA, realizing a versatile test setup for cQED experiments at high magnetic field with a quantum-limited amplifier. 

\begin{acknowledgements}
We thank Timur Zent for making the CAD drawings of the magnetic shields and helping with the assembly of the cryogenic setup. 
We thank Arno Bargerbos for his advice and for sharing his experience with TWPAs in magnetic shields. 
We thank Tom Paquin from  MuShield for competent advice on magnetic shields as well as speedy delivery.
We thank Daniel Strange and Riccardo Vianna from Oxford Instruments for sharing detailed data on the stray field of our magnet. 
This project has received funding from the European Research Council (ERC) under the European Union’s Horizon 2020 research and innovation program (grant agreement No 741121) and was also funded by the Deutsche Forschungsgemeinschaft (DFG, German Research Foundation) under CRC 1238 - 277146847 (Subproject B01) as well as under Germany’s Excellence Strategy - Cluster of Excellence Matter and Light for Quantum Computing (ML4Q) EXC 2004/1 - 390534769. 
\end{acknowledgements}

\section*{author contributions}

The project was conceived by C.D. and Y.A. in coordination with L.P..
The TWPAs were fabricated by A.C. and L.P. and factory characterized by G.B., A.C., and L.P..
The measurements were performed by L.M.J. and C.D. with help from J.K. and advice from G.B., L.P.. 
L.M.J. and C.D. built the setup for the TWPA in the fridge with the magnet with magnetic shields and bypass switches. 
The data was analyzed by L.M.J. and C.D. with help from G.B., N.R., and G.C..
G.C. provided theory support for the project.
The manuscript was written by L.M.J. and C.D. with input from all coauthors.

\section*{Software}

The setup was controlled based on QCoDeS (https://github.com/QCoDeS/Qcodes) drivers and logging~\cite{Qcodes}, 
The measurements were run using Quantify-core (https://gitlab.com/quantify-os/quantify-core)~\cite{rol2021quantify}.

\section*{Data availability}

Datasets and analysis in the form of Jupyter notebooks that create the figures are available on Zenodo with doi \href{https://doi.org/10.5281/zenodo.10728283}{10.5281/zenodo.10728283}.

\appendix

\section{Device descriptions}
\label{sec:device_description}

\begin{table*}
\centering
\begin{tabular}{c | c c c c c c c c c c c} 
 \hline
 Device & $L_\mathrm{J}$ & $C_\mathrm{g}$  & $C_\mathrm{J}$ & $w$  & $\bar{h}$ & $\eta$ & $N_\mathrm{J}$ & $N\mathrm{p}$ & $f_\mathrm{g}$ &  \makecell{mean SNR \\improvement} & \makecell{electrical \\ losses } \\ 
 [1ex]
  &  [\SI{}{\pico\henry}] &[\SI{}{\femto\farad}] & [\SI{}{\femto\farad}]& [\SI{}{\micro\meter}] & [\SI{}{\micro\meter}]& & & & [\SI{}{\giga\hertz}]&   [\SI{}{\decibel}] & [\SI{}{\decibel}] \\  [1ex]
 
 \hline
 TWPA A &  95 & 38 & 500 & 0.7 & 16 & $\pm$\SI{5}{\percent} & 1596 & 28 & 8.7 & 6 & 4-6 \\ [1ex]
 TWPA B & 133 & 29 & 490 & 0.7 & 16 & $\pm$\SI{5}{\percent} & 1800 & 33 & 7.2 & 7 & 3-6 \\ [1ex]
 \hline
\end{tabular}
\caption{
Parameters and factory characterization figures of merit for the two TWPAs in this work. Parameters are used in the modeling. 
}
\label{tab:twpa_params}
\end{table*}

The TWPAs in this work are of the Argo model, commercially available and made by Silent Waves.
The field dependence with the TWPA at the magnet center is measured on TWPA A, while the shield performance and the temperature dependence are measured on TWPA B.
The TWPAs are fabricated and packaged by Silent Waves and delivered as a module with SMA connectors.
The two TWPAs are similar to the one described in Ref.~\onlinecite{Planat2020} but have single-JJs instead of SQUIDs.
The JJs are fabricated using double-angle evaporation with a bridge-free design~\cite{Lecocq2011}.
The film thicknesses of the first and second aluminum layer are \SI{20}{\nano\meter} and \SI{50}{\nano\meter} respectively.
The ground plane is made of a thick copper layer, so there are no large extended aluminum areas.
Detailed parameters for both devices can be found in \cref{tab:twpa_params}.

Silent waves measured the TWPA figures of merit in their setup before shipping.
Electrical losses in the TWPA, a combination of insertion loss and loss within the TWPA, were estimated to be 4-\SI{5}{\decibel} in the factory calibration. 
These losses have to be offset by the gain to improve performance.
The factory calibration estimated an SNR improvement of \SI{6}{\decibel} for a gain similar to the one measured in this work.

\section{Measurement setup for magnetic field dependence}
\label{sec:measurement_setup}

\begin{figure}
  \centering
  \includegraphics[width=\columnwidth]{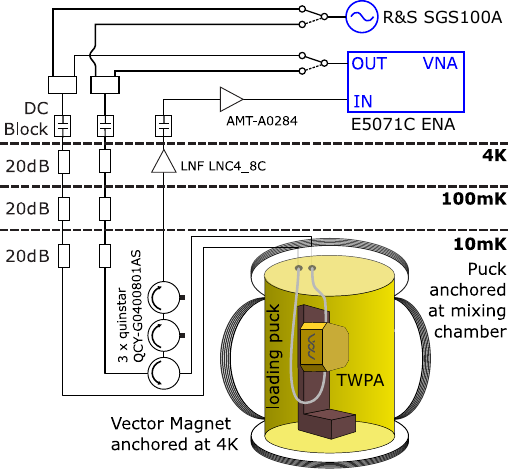}
%  \end{center}
  \caption{
    Wiring diagram and setup for the measurements with the TWPA inside the puck.
    The TWPA is mounted roughly at the center of the vector magnet and the field dependence is studied. 
    Both reflection and transmission measurements were possible.
    }
  \label{fig:wiring_diagram}
\end{figure}

The measurement setup for the magnetic field dependence of the TWPA is illustrated in \cref{fig:wiring_diagram}.
The TWPA experiments were performed with the device in the puck of a bottom-loading dilution refrigerator (Triton 500, Oxford Instruments) with a nominal base temperature of $\sim$\SI{10}{\milli\kelvin}. 
A 3-axis \SI{6}{\tesla}/\SI{1}{\tesla}/\SI{1}{\tesla} vector magnet is used to apply magnetic fields to the sample.
In this paper, we do not perform field cooling of the TWPA, but we change the magnetic field while keeping the device at base temperature. 
During the cooldown, the magnet was shorted and disconnected from its power supply to avoid the creation of vortices during the superconducting transition of the TWPA. 
For the $B_{\parallel,1}$ axis we mainly used the Oxford Instruments Mercury iPS power supply, which can supply sufficient current to reach a magnetic field of \SI{1}{\tesla}.
For the other two axes, we used a Keithley 2461 SourceMeter as a current source which can only reach magnetic fields of \SI{0.16}{\tesla} but is more fine-grained and less noisy.
For all measurements of the shielded setup, the Mercury iPS was used for all field directions, so we could see the shield performance with up to \SI{6}{\tesla} of magnetic field at the sample.

The alignment of the magnet axes was such that $B_\mathrm{x}$, $B_\mathrm{y}$, $B_\mathrm{z}$, correspond to $B_\perp$, $B_{\parallel,1}$, $B_{\parallel,2}$.
We took angular scans of the magnetic field at fixed $\left \vert B \right \vert$ to estimate the degree of misalignment, but in particular for the $B_\perp$ direction, the hysteresis is more significant than any potential misalignment.
Misalignment is likely less than \SI{1}{\degree}.

The TWPA transmission and gain measurements were performed with a Vector Network Analyzer, and an additional microwave source was used as a pump.
The wiring was such that we could use a room temperature microwave switch to measure the TWPA in reflection or transmission and to also reverse the pump direction (see wiring diagram \cref{fig:wiring_diagram}).
We use 60 dB of attenuation in the input lines to ensure that the TWPA input noise is low, but we do not use Eccosorb or other filters here that would avoid pair-breaking radiation.
For operating the TWPA in an actual experiment, a directional coupler would be used to couple in the pump tone, but to just measure the TWPA itself this is not necessary. 
The fridge wiring is not entirely optimal for the TPWA, as the bandgap is slightly above \SI{8}{\giga\hertz}, outside the specifications of the circulators that were mounted in the fridge.
For the mounting of the TWPA in the fridge with shields, wider range circulators were used (see \cref{sec:TWPA_in_fridge_setup}).

\section{TWPA model}
\label{sec:twpa_model}

Here we present the model for the TWPA with the modulation in the junction dimension $h$, see Fig.~\ref{fig:fig1}~(b). The model is based on that of Ref.~\cite{Planat2020} and is used to derive formulas for the plasma and bandgap frequencies which can be directly compared with experiment. The model also returns the effective impedance of the array, which can be used to understand the reflection at the input and output.
In the first part of this appendix, we focus on the effect of field $B_{\parallel,1}$. The other in-plane field direction is treated in subsection~\ref{app:Bpar2}. Subsection~\ref{sec:Delta_vs_B} concerns the suppression of the superconducting gap by the field, and subsection~\ref{sec:Bcrit_par_vs_perp} considers the perpendicular field.

The phase-matching of the TWPA is achieved by periodic modulation of the Josephson inductance and capacitance, by varying the size of the junctions. 
Let $N_\mathrm{J}$ be the number of junctions in the TWPA. 
Then we have $N_\mathrm{J} + 1$ superconducting islands, which we will index with integers $n = 0,1,...,N_\mathrm{J}$. 
We will index the Josephson junction between two islands with half-integers: the junction between islands $n$ and $n+1$ is indexed $n+1/2$.
Since the modulation of the Josephson junctions is periodic, we can define the entire array using a small set of parameters. 
Let $\bar{L}_\mathrm{J}$ and $\bar{C}_\mathrm{J}$ be the average Josephson inductance and capacitance respectively, and let $\bar{C}_\mathrm{g}$ be the average capacitance to ground of an island. 
Define $N_\mathrm{p}$ the period of the modulation, $\eta$ the junction area modulation amplitude. Then we have for any island $n$ and any junction $n+1/2$ that 
\begin{align}
    L_{\mathrm{J},n+1/2}^{-1} & = \bar{L}_\mathrm{J}^{-1}[1+\eta \cos(G (n+1/2))], \label{eq:LJn} \\ 
    C_{\mathrm{J},n+1/2} & = \bar{C}_\mathrm{J}[1+\eta \cos(G (n+1/2))], \label{eq:CJn} \\ 
    C_\mathrm{g,n} & = \bar{C}_\mathrm{g}[1+\eta /2 \sum_{ \pm } \cos(G (n \pm 1/2))],
\end{align}
where $G=2\pi/N_\mathrm{p}$.
Note that the modulation amplitude of the junction area determines the modulation of $L_\mathrm{J}$, $C_\mathrm{J}$ and $C_\mathrm{g}$.
For $L_\mathrm{J}$ and $C_\mathrm{J}$, this is clear: these are inversely proportional and proportional to the junction area, respectively. The ground capacitance
$C_\mathrm{g}$ is proportional to the area of the islands because it is achieved using a single ground plane on top of the entire junction array.
Then the modulation of $C_\mathrm{g}$ follows from the geometry of the device, as
each island consists of two junction areas, and a negligible area connecting the two junction areas  (see \cref{fig:fig1} \textbf{(b)}).
Hence the modulation amplitude of the area of an island is the same as that of the junctions, but we need to average the modulation of two junctions.

We first consider the plasma frequency $f_\mathrm{p}$, above which transmission through the TWPA is strongly suppressed.
For any junction it is given by $ \omega_\mathrm{p} = 1/\sqrt{L_{\mathrm{J},n+1/2}C_{\mathrm{J},n+1/2}}$.
Note that the modulation cancels so that at zero field we get the same plasma frequency for each junction:  
\begin{equation}
\omega_\mathrm{p}(0) = 1 / \sqrt{ \bar{L}_\mathrm{J}  \bar{C}_\mathrm{J}}.
\end{equation}
In general the Josephson inductance $L_\mathrm{J}$ is related to the critical current $I_\mathrm{c}$ of a JJ via $\bar{L}_\mathrm{J} = \Phi_\mathrm{0}/(2\pi I_\mathrm{c})$, where $\Phi_\mathrm{0}$ is the superconducting flux quantum and $I_\mathrm{c}$ is proportional to the superconducting gap $\Delta$ and the junction's conductance.
An in-plane magnetic field is perpendicular to the current flow in the tunnel barrier, so we expect a Fraunhofer pattern in the critical current $I_\mathrm{c}$. 
Additionally, the superconducting gap is suppressed with field, causing a decrease in $I_\mathrm{c}$ (the gap suppression is discussed in more detail in Appendix~\ref{sec:Delta_vs_B}). 
Therefore, we model the in-plane field dependence of the critical current as
\begin{equation}\label{eq:IcB}
 I_\mathrm{c}(B) = I_\mathrm{c,0} \left \vert \sinc(\pi\phi/\Phi_\mathrm{0}) \right \vert \left[\Delta(B/B_\mathrm{c}) / \Delta_\mathrm{0}\right]
\end{equation}
where $I_\mathrm{c,0}$ and $\Delta_\mathrm{0}$ denote quantities in zero field, $\phi=B A$ with $A$ the area penetrated by flux [cf. Fig.~\ref{fig:fig1}~(b)], $\mathrm{sinc}(x)=\sin(x)/x$, and $B_\mathrm{c}$ is an appropriately chosen critical field.

\subsection{$B_{\parallel,1}$ direction}
\label{app:Bpar1}

For the $B_{\parallel,1}$ magnetic field direction, the Fraunhofer-type modulation of $I_\mathrm{c}$ is identical for all junctions in the array because the field is applied perpendicular to $w$, the dimension of the Josephson junction that is not modulated.
As such, we can apply Eq.~\eqref{eq:IcB} to $\bar{L}_\mathrm{J}$, rather than separately to each junction.
Therefore
\begin{equation}\label{eq:plasmaB1}
    \omega_\mathrm{p} \left(B_{\parallel,1}\right) = \omega_\mathrm{p}(0) \sqrt{\left|\mathrm{sinc}\left(\pi\frac{B_\mathrm{\parallel,1}}{B_\mathrm{\Phi,1}}\right)\right|\frac{\Delta(B_{\parallel,1}/B_\mathrm{c})}{\Delta_0}}.
\end{equation}
with $B_{\Phi,1}=\Phi_0/(l w)$.

The Fraunhofer pattern depends on junction geometry as well as on the spatial distribution of the current through the junction. 
In particular, the sinc functional dependence is derived for a rectangular junction with uniform current. 
In Ref.~\onlinecite{Barone1977}, a one-parameter phenomenological model was proposed (and compared to experiment) for a rectangular junction that can interpolate between uniform current and current concentrated at the edges of the junction. 
We assume the current density profile 
\begin{equation}\label{eq:currentdensityprofile}
    J = J_\mathrm{max} \frac{\mathrm{cosh}(2 x \chi/L )}{\mathrm{cosh}(\chi)},
\end{equation}
where $L$ is the size of the junction (either $h$ or $w$), and $x \in [-0.5L,0.5L]$.
The uniform current case is recovered in the limit $\chi \to 0$.
For increasing $\chi$ the current density in the interior decreases while the edges stay at $J_\mathrm{max}$.
The corresponding Fraunhofer pattern amounts to the substitution in Eq.~\eqref{eq:plasmaB1}
\begin{equation}\label{eq:sincreplacement}
    \mathrm{sinc}\, y \to \mathcal{F}(y,\chi) \equiv \frac{\chi^2}{\chi^2+y^2}\left[\frac{y \sin y}{\chi \tanh \chi} + \cos y\right].
\end{equation}

We now turn to the calculation of the bandgap in the low-power limit. In this regime
to calculate the edges of the bandgap, we can use the linearized version of a set of equations found in Appendix~B of Ref.~\cite{Planat2020}:
\begin{align}
    & \left[\frac{\omega^2}{\omega_\mathrm{p}^2 \ell_\mathrm{s}^2}-k^2 \left(1-\frac{\omega^2}{\omega_\mathrm{p}^2}\right)\right] \mathrm{A}  \\ & + \left[\frac{\eta}{2}\frac{\omega^2}{\omega_\mathrm{p}^2 \ell_\mathrm{s}^2} +\frac{\eta}{2}k(G-k) \left(1-\frac{\omega^2}{\omega_\mathrm{p}^2}\right)\right] \mathrm{B}=0, \nonumber \\
    & \left[\frac{\eta}{2}\frac{\omega^2}{\omega_\mathrm{p}^2 \ell_\mathrm{s}^2} +\frac{\eta}{2}k(G-k) \left(1-\frac{\omega^2}{\omega_\mathrm{p}^2}\right)\right] \mathrm{A}  \\ & + \left[\frac{\omega^2}{\omega_\mathrm{p}^2 \ell_\mathrm{s}^2}-(G-k)^2 \left(1-\frac{\omega^2}{\omega_\mathrm{p}^2}\right)\right] \mathrm{B} = 0. \nonumber
\end{align}
relating the amplitudes $\mathrm{A}$ and $\mathrm{B}$ of waves with wavevector $k$ and $k-G$ and frequency $\omega$.
Here $\ell_\mathrm{s}$ is the Coulomb screening length, $\ell_\mathrm{s} = \sqrt{\bar{C}_\mathrm{J} / \bar{C}_\mathrm{g}}$.
For $\eta=0$ (no modulation) the two equations are decoupled and give degenerate solutions for $k=G/2$ (that is, at the band edge), and for $\eta > 0$ a bandgap opens at $k=G/2$ (see for example Fig.~1 in Ref.~\cite{Planat2020}).
Therefore, by finding the frequencies for which the wave with $k=G/2$ propagates, we find the frequencies of the edges of the bandgap.
Setting $k=G/2$ and the determinant of the system to zero we find the values of $\omega$ for which this system has a non-trivial solution:
\begin{equation}\label{eq:bandgap1}
    \omega_\mathrm{g,\pm} = \omega_\mathrm{p}\sqrt{\frac{(G/2)^2(1\pm\eta/2)}{(G/2)^2(1\pm\eta/2)+1/\ell_\mathrm{s}^2 (1\mp\eta/2)}}
\end{equation}
where the prefactor $\omega_\mathrm{p}$ on the right-hand side should be understood to depend on magnetic field as in Eq.~\eqref{eq:plasmaB1}.

With this formula, we can fit the experimental data. As shown in \cref{fig:fig2}, the center of the gap is extracted for each field value where the gap is visible. We can take the average of the frequencies found with Eq.~\eqref{eq:bandgap1} and run a least-squares fitting procedure, where we vary $\omega_\mathrm{p} (0)$, $\chi$ and $B_{\Phi,1}$. For the other parameters, we use the designed values of the TWPA A (see \cref{tab:twpa_params}).
With this model, we can not fit $\bar{C}_\mathrm{J}$ and $\bar{C}_\mathrm{g}$ independently, since they enter into the formula for bandgap through their ratio (in $\ell_\mathrm{s}$) and through $\omega_\mathrm{p}(0)$ (which depends also on the Josephson inductance).
The critical field $B_\mathrm{c}$ can be estimated directly from the data, considering the value at which transmission becomes undetectable, so we use the value $B_\mathrm{c} =  \SI{236}{\milli\tesla}$. Now we can fit the parameters mentioned: $$f_\mathrm{p} (0) = \SI{23.0}{\giga\hertz}, \, \chi = 0.668, \, B_{\Phi,1} = \SI{107.8}{\milli\tesla}.$$

\subsection{$B_{\parallel,2}$ direction}
\label{app:Bpar2}

For the $B_{\parallel,2}$ direction, due to the modulation in the dimension $h$ of the junctions normal to the field [cf. Fig.~\ref{fig:fig1} (b)], we have to modify the above results to take into account that the Fraunhofer effect modulates the Josephson inductance of the different junctions by different amounts.
Considering junction $n+1/2$, its inductance and capacitance are [cf. Eqs.~\eqref{eq:LJn} and \eqref{eq:CJn}] 
\begin{align}
    L_{J, n+1/2}^{-1} & = \bar{L}^{-1}\left\{1+\eta \cos\left[G(n+1/2)\right]\right\}\left|\mathrm{sinc}\left(\pi\frac{B_{\parallel,2}}{B_{\Phi,2}^{(n)}}\right)\right| \label{eq:Lmod}, \\
    C_{J, n+1/2} & = \bar{C}\left\{1+\eta \cos\left[G(n+1/2)\right]\right\},
\end{align}
with
\begin{equation}
    B_{\Phi,2}^{(j)} = \bar{B}_{\Phi,2}/\left\{1+\eta \cos\left[G(n+1/2)\right]\right\}
\end{equation}
where, based on the junction geometry (see Table~\ref{tab:twpa_params}), we can relate $\bar{B}_{\Phi,2}$ to $B_{\Phi,1}$:
\begin{equation}\label{eq:Bbp1}
    \bar{B}_{\Phi,2} = (0.7/16) B_{\Phi,1}
\end{equation}
In practice, we treat $B_{\Phi,2}$ as a fit parameter and then check for consistency with the expectation from this equation; we find a discrepancy of order 4~\%, see Sec.~\ref{sec:twpa_vs_Bpar2_Bperp}. 
This small difference might effectively account for the possibly different current distributions in the two directions, which are not known (in other words, the $\chi$ values, which for simplicity we assume to be identical, could, in fact, differ for the two directions).

In writing Eq.~\eqref{eq:Lmod} we assume $B_{\parallel,2} \ll B_\mathrm{c}$ so we can ignore gap suppression for simplicity; it can be straightforwardly included by multiplying the right-hand side of Eq.~\eqref{eq:Lmod} by $\Delta(B_{\parallel,2}/B_\mathrm{c})/\Delta_0$, see Eq.~\eqref{eq:IcB}.
The measurable plasma frequency is then given by the smallest one at the given field:
\begin{align}\label{eq:plasmaB2}
    & \omega_\mathrm{p} \left(B_{\parallel,2}\right) = \min_{\{n\}} \left[ \frac{1}{\sqrt{L_{J, n+1/2}C_{J, n+1/2}}}\right] \\& = \min_{\{n\}} \left[ \omega_\mathrm{p}(0) \sqrt{\left|\mathrm{sinc}\left(\pi\frac{B_{\parallel,2}}{B_{\Phi,2}^{(n)}}\right)\right|\frac{\Delta(B_{\parallel,2}/B_\mathrm{c})}{\Delta_0}} \right] \nonumber
\end{align}

For the bandgap, assuming $\eta \ll 1$ and $\eta \pi B_{\parallel,2}/\bar{B}_\Phi \lesssim 1$, we can find a simple generalization of Eq.~\eqref{eq:bandgap1} as follows: we expand Eq.~\eqref{eq:Lmod} to first order in $\eta$ and cast the result in the form similar to Eq.~\eqref{eq:Lmod} itself,
\begin{equation}
    L_{J, n+1/2}^{-1} \simeq \bar{L}^{-1}\left\{1+\beta\eta \cos\left[G(n+1/2)\right]\right\}\left|\mathrm{sinc}\left(\pi\frac{B_{\parallel,2}}{\bar{B}_{\Phi,2}}\right)\right|
\end{equation}
with the field-dependent factor
\begin{equation}\label{eq:betadef}
    \beta = \pi\frac{B_{\parallel,2}}{\bar{B}_{\Phi,2}} \cot \left(\pi\frac{B_{\parallel,2}}{\bar{B}_{\Phi,2}}\right)
\end{equation}
or, if we use the current distribution assumption leading to the substitution in Eq.~\eqref{eq:sincreplacement},
\begin{align}
    &\beta =  \\
    &\frac{\frac{y \sin y}{\chi \tanh \chi}\left(y \cot y +\frac{2\chi^2}{\chi^2+y^2}\right)+\cos y \left(1-y \tan y - \frac{2y^2}{\chi^2+y^2}\right)}{\frac{y \sin y}{\chi \tanh \chi} + \cos y}, \nonumber
\end{align}
where $y = \pi B_{\parallel,2}/\bar{B}_{\Phi,2}$.
Then repeating the procedure used to arrive at Eq.~\eqref{eq:bandgap1}, we find that the expression for the bandgap has the same form as in that equation if we replace $\eta \to \beta\eta$ in the numerator and
\begin{equation}
    \omega_\mathrm{p} \to \omega_\mathrm{p}(0) \sqrt{\left|\mathrm{sinc}\left(\pi\frac{B_{\parallel,2}}{\bar{B}_{\Phi,2}}\right)\right|\frac{\Delta(B_{\parallel,2}/B_\mathrm{c})}{\Delta_0}}
\end{equation}
in the prefactor.

The field values $B_{\parallel,2}^{(n)}$ at which the bandgap closes can be found by requiring $\omega_+=\omega_-$, which leads to the condition
\begin{equation}\label{eq:bandclose}
    \beta = \frac{(G/2)^2-1/\ell_\mathrm{s}^2}{(G/2)^2+1/\ell_\mathrm{s}^2} \equiv \beta_\mathrm{c}
\end{equation}
This should be contrasted with the condition $\beta=0$ that corresponds to the (approximate) suppression of the spatial modulation of the Josephson inductance (or critical current). However, with our parameters we have $\beta_\mathrm{c}\simeq -0.716$ and an approximate solution to Eq.~\eqref{eq:bandclose}, 
when using Eq.~\eqref{eq:betadef} for $\beta$, is
\begin{equation}
    B_{\parallel,2}^{(n)} \simeq \bar{B}_{\Phi,2} \left[\left(n+\frac12\right) -\frac{\beta_\mathrm{c}}{\pi^2(n+1/2)}\right], \quad n=0,\,1,\, \ldots
\end{equation}
The first term on the right-hand side satisfies the condition $\beta=0$; the correction -- that is, the second term -- becomes smaller (and the approximation more accurate) as $n$ increases.

Finally, we note that an approximate position for the center $\omega_c$ of the bandgap is found by setting $\eta=0$ in Eq.~\eqref{eq:bandgap1}, $\omega_c = \omega_\mathrm{p} G/2\sqrt{(G/2)^2+1/\ell_s^2}$. This formula can be easily generalized to estimate the position of the next bandgap by replacing $G\to 2G$, cf. Fig.~\ref{fig:fig4}~(a).

\subsection{Field dependence of the superconducting gap for in-plane field}
\label{sec:Delta_vs_B}

Here we consider the dependence of the gap on the parallel magnetic field.
Such a field can suppress superconductivity through two mechanisms: the Zeeman splitting of the electrons forming a Cooper pair, and the so-called orbital effect -- the suppression of the order parameter by the supercurrent that arises to screen the applied magnetic field.
In films thin compared to the penetration depth, $t\ll \lambda$, the screening of the field can be ignored, but not the effect of the supercurrent on the order parameter.
Whether the Zeeman effect plays a significant role or not depends on the dimensionless parameter 
\begin{equation}
    c =\frac{D (et)^2\Delta_0}{6\hbar \mu_\mathrm{B}^2} f\left(\ell/t\right)
\end{equation}
where $D$ is the electron diffusion constant, $\ell$ the mean-free path, and $\mu_\mathrm{B}$ the Bohr magneton~\cite{Maki2018}.
The dimensionless function in the last factor has the limits $f(0) = 1$ and $f(x)\simeq 3/4x$ for $x\gg1$.
When $c>1$, the order parameter suppression is dominated by the orbital effect and the Zeeman splitting can be ignored.
Experimentally, it was found that $c\sim 1$ for films with $t\sim 7$-8\,nm~\cite{Catelani2008,Adams2017}.
Given the rapid increase of $c$ with thickness, even for the thinnest film forming the Josephson junctions, $t\simeq 20\,$nm, the Zeeman effect plays only a negligible role.
On the other hand, we note that the thicker film, $t\simeq 50\,$nm, is still thinner than the penetration depth in Al films of this thickness, $\lambda > 100\,$nm~\cite{Lopez2023}.

\begin{figure}
    \centering
    \includegraphics[width=\columnwidth]{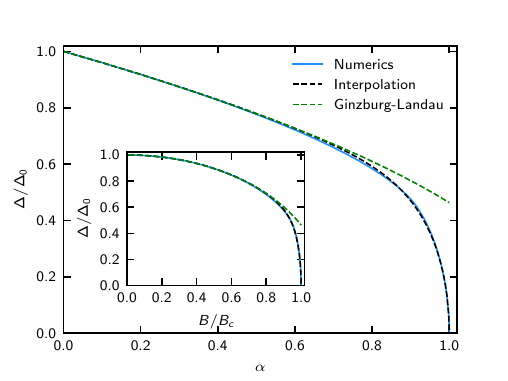}
    \caption{Normalized order parameter as function of the pair-breaking parameter $\alpha = (B/B_\mathrm{c})^2$. Solid blue line: numerical solution to the equation of Ref.~\cite{Skalski1964} determining the gap suppression at zero temperature within AG theory. 
    Black, dashed: analytical interpolation, Eq.~\eqref{eq:Da}. Green, dashed: GL type approximation, cf. Eq.~\eqref{eq:GL}. Inset: same as the main panel, but plotted as a function of the parallel magnetic field normalized by the critical one.
    }
    \label{fig:interpolation}

\end{figure}

Having to consider only the orbital effect simplifies the calculation of the order parameter suppression, which is analogous to that due to paramagnetic impurities, as studied by Abrikosov and Gorkov (AG)~\cite{Abrikosov1961, Skalski1964}.
At low temperatures $T\ll T_\mathrm{c}$, we can use the $T=0$ expressions up to exponentially small corrections that we ignore. The amount of gap suppression is determined by a dimensionless pair-breaking parameter $\alpha$ which in the present case can be written as $\alpha=(B/B_\mathrm{c})^2$, where $B_\mathrm{c}$ is the parallel critical field (in the notation of Ref.~\cite{Skalski1964} we have the relation $\Gamma=\alpha\Delta_0/2$, where $\Gamma$ is their dimensionful pair-breaking parameter). While the implicit transcendental equation determining $\Delta$ as a function of $\alpha$ has a relatively simple form amenable to straightforward numerical solution [see Eq.~(3.5) in Ref.~\cite{Skalski1964}], no closed form solution is available. Here we propose a formula that interpolates between the analytically tractable limits $\alpha \to 0$ and $\alpha \to 1$, the latter value representing the critical one at which superconductivity disappears:
\begin{equation}\label{eq:Da}
    \Delta(\alpha) = \Delta_0 \sqrt{1-\frac{\pi}{4}\alpha - \left(1-\frac{\pi}{4}\right)\alpha^\gamma}
\end{equation}
with $\gamma = (12-\pi)/(4-\pi) \simeq 10.32$. 
As shown in Fig.~\ref{fig:interpolation} this formula is a good approximation to the numerically exact solution, deviating less than 2\% from the latter.

Note that due to the large exponent $\gamma$, the last term under the square root in Eq.~\eqref{eq:Da} can be dropped if $\alpha$ is not too close to unity; even with this additional approximation the deviation from numerics remains below 2\% for $\alpha\lesssim 0.7$, corresponding to $B/B_\mathrm{c} \lesssim 0.84$ (see inset). In this range, we can therefore write
\begin{equation}\label{eq:GL}
    \Delta(B/B_\mathrm{c}) \simeq \Delta_0 \sqrt{1-\left(\frac{B}{\tilde{B}_\mathrm{c}}\right)^2}
\end{equation}
where $\tilde{B}_\mathrm{c} = 2 B_\mathrm{c}/\sqrt{\pi} \simeq 1.128 B_\mathrm{c}$. This shows that up to fields not too close to the critical one, the gap suppression (at zero temperature) takes the form familiar from Ginzburg-Landau theory [valid near the critical temperature, see Eq.~(4.52) in Ref.~\cite{Tinkham04}], although the effective critical field $\tilde{B}_\mathrm{c}$ entering this expression overestimates the actual critical field $B_\mathrm{c}$ by almost 13\%.

\begin{figure*}
  \centering
      \includegraphics[width=\textwidth]{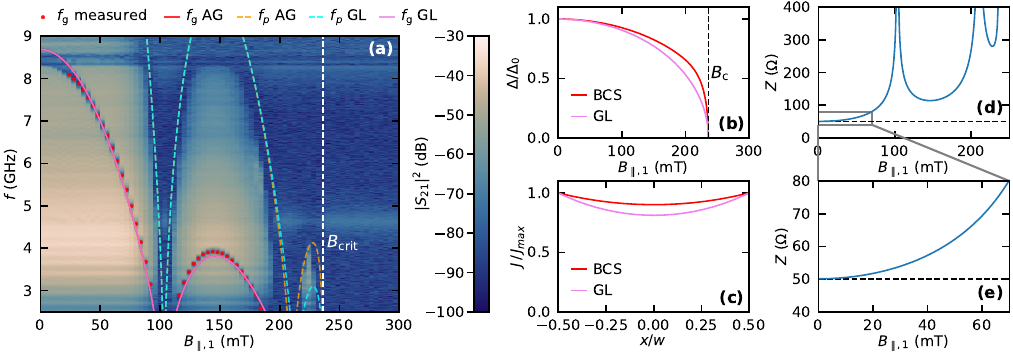}
%  \end{center}
  \caption{
  Details on the $B_{\parallel,1}$ dependence of the pump-off TWPA transmission.
  \textbf{(a)} TWPA transmission spectrum without pump tone as a function $B_{\parallel,1}$, same as in~\cref{fig:fig2} \textbf{(a)}. It shows the same fit for $f_{g}$ and $f_{p}$, and an additional fit using Ginzburg-Landau dependence for $\Delta$. 
  The fits overlap for low fields because away from $B_\mathrm{c}$ the difference in $\Delta(B)$ [see \textbf{(b)}] can be compensated by adapting the current distribution in the model  [see \textbf{(c)}]; however, only the AG fit can accurately capture the second Fraunhofer lobe. 
  \textbf{(d)} and \textbf{(e)} Impedance of the TWPA junction array as a function of field. While the junction array is impedance-matched to the environment at zero field, mismatch increases as the $f_\mathrm{g}$ is tuned lower. 
  This partially explains the reduction in gain with increasing field; in particular, in the higher Fraunhofer lobes, there is always a considerable impedance mismatch. 
  }
  \label{fig:details_Bpar1_vs_B}
\end{figure*}

Given the above discussion of the approximate expression in Eq.~\eqref{eq:GL}, one may wonder if the data can be satisfactorily fitted just by using the Ginzburg-Landau formula shown there.
However, setting the critical field to the measured value cannot reproduce the $f_\mathrm{p}$ of the second Fraunhofer lobe accurately, see \cref{fig:details_Bpar1_vs_B}~\textbf{(a)}.
The thin-film AG gap dependence fits $f_\mathrm{p}$ more accurately because $\Delta$ remains higher and then is more quickly suppressed when increasing the field towards the critical one, as seen in \cref{fig:details_Bpar1_vs_B}~\textbf{(b)}. On the other hand, the bandgap frequency
$f_\mathrm{g}$ can be fitted as accurately with the GL formula as with the AG one because we cannot measure it as close to $B_\mathrm{c}$ and we can use the additional fit parameter $\chi$ to adjust the curve.
In both cases, we need to assume a higher current density at the edges of the JJs, but to perform the fitting with the GL expression we need to assume a less uniform current density than for the AG case, see \cref{fig:details_Bpar1_vs_B}~\textbf{(c)}.

Based on the model we can also estimate the TWPA impedance $Z$ as a function of field, see \cref{fig:details_Bpar1_vs_B} \textbf{(d)} and \textbf{(e)}.
The impedance of the array is equal to $Z=\sqrt{\bar{L}_\mathrm{J}/\bar{C}_\mathrm{g}}$.
The capacitance to ground $\bar{C}_\mathrm{g}$ does not change with magnetic field, and we obtain a value for $\bar{L}_\mathrm{J}$ at any magnetic field from the fit. Therefore we can calculate the $Z$ at any value of $B_{\parallel,1}$.
At zero field we find that the TWPA is well matched to \SI{50}{\ohm}, but the tuning of $f_\mathrm{g}$ with $B_{\parallel,1}$ is accompanied by an increasing impedance, which diverges as the field approaches a Fraunhofer minimum.
Therefore, the insertion loss at the TWPA grows, showing that tunability by field in the TWPA would always come at the cost of some reduction in performance.

\subsection{$B_\perp$ direction}
\label{sec:Bcrit_par_vs_perp}

Here we briefly discuss some aspects of the TWPA behavior in perpendicular field.
In \cref{fig:fig4} \textbf{(c)} and \textbf{(d)} we see that the field where the breakdown in $\vert S_{21}\vert$ occurs varies by scan direction, but $f_\mathrm{g}$ is also tuned down with $B_\perp$.
We can try to estimate $B_{\mathrm{c},\perp}$ by fitting the $f_\mathrm{g}$ dependence.
The dependence of $f_\mathrm{g}$, suggests a GL-like expression [cf. Eq.~\eqref{eq:GL}] of the form 
\begin{equation}
f_\mathrm{g}(B) = f_\mathrm{g}(0) \sqrt{1-\left (\frac{B_\perp-B_\mathrm{offset}}{B_{\mathrm{c},\perp}} \right)^2} 
\end{equation}
for the bandgap edges with a different $B_\mathrm{offset}$ for the two sweep directions.
We do not consider a Fraunhofer contribution in this direction as the field is parallel to the JJ current.
This model can describe the data reasonably well for a plausible $B_{\mathrm{c},\perp}=\SI{10.3}{\milli\tesla}$, which is also consistent with the breakdown in $\vert S_{21} \vert$ [vertical dashed lines in Figs.~\ref{fig:fig4} (b) and (c)]. 
The critical field $B_{\mathrm{c},\perp}$ can be related to the coherence length $\xi$ by~\cite{Tinkham04}
\begin{equation}
 B_{\mathrm{c},\perp} = \frac{\Phi_0}{2\pi\xi^2},
\end{equation}
which gives $\xi\simeq\SI{180}{\nano\meter}$. 
Also, the ratio between $ B_{\mathrm{c},\perp}$ and the parallel critical field $B_{\mathrm{c},\parallel}$ for thin films depends on film thickness $t$ as~\cite{Tinkham04}
\begin{equation}
    \frac{B_{\mathrm{c},\perp}}{B_{\mathrm{c},\parallel}} = \frac{t}{2\sqrt{3}\xi}.
\end{equation}
From this expression and the measured critical fields, we would estimate $t\simeq\SI{27}{\nano\meter}$, within a factor of 2 from the thickness of the thicker film (which has a lower critical field compared to the thinner one).

Finally, let us consider the question of hysteresis, which we assume is due to the presence of at least one stable vortex in some of the junctions. 
For a superconducting strip of width $w$, the field $B_\mathrm{L}$ at which a vortex is stable is given by~\cite{Kuit2008}
\begin{equation}
    B_\mathrm{L} = \frac{2\Phi_0}{\pi w^2} \ln \left(\frac{2w}{\pi \xi}\right) \simeq 2.4\,\mathrm{mT}
\end{equation}
where we used the width $w$ in \cref{tab:twpa_params} and the coherence length estimated above from the perpendicular critical field. The value is in agreement with the observed \SI{2.4}{\milli\tesla} hysteresis.

\section{TWPA transmission and reflection}
\label{sec:TWPA_transmission_vs_reflection}

\begin{figure}
  \centering
  \includegraphics[width=\columnwidth]{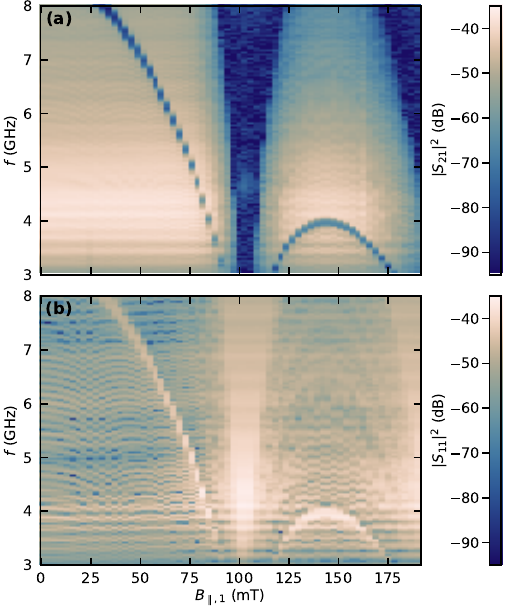}
%  \end{center}
  \caption{
    TWPA transmission \textbf{(a)} and reflection \textbf{(b)} vs $B_{\parallel,1}$.
    It is apparent that when the TWPA transmission is reduced at the Fraunhofer minimum, the reflection is enhanced and reaches similar levels.
    }
  \label{fig:TWPA_transmission_vs_reflection}
\end{figure}

As mentioned in \cref{sec:measurement_setup} the setup allowed measuring the TWPA transmission $\vert S_{21}\vert$ and the reflection $\vert S_{11}\vert$ using a circulator at the TWPA output.
In the modeling above, we focus on the frequencies of TWPA features ($f_\mathrm{g}$ and $f_\mathrm{p}$), and in the gain measurements, the calibration of the fridge lines is not so essential. 
This is partly because the setup to measure the TWPA in field was not particularly adapted to TWPA characterization experiments, where one would carefully calibrate the input and output lines to understand the insertion loss and dissipation inside the TWPA.
Nonetheless, we have measured the TWPA $\vert S_{21}\vert$ and $\vert S_{11}\vert$, see \cref{fig:TWPA_transmission_vs_reflection} \textbf{(a)} and \textbf{(b)}.
The transmission and reflection input have nominally identical attenuation and the electrical losses in the TWPA are on the order of \SI{4}{\decibel}, meaning one can roughly compare $\vert S_{21}\vert$ and $\vert S_{11}\vert$.
We see that close to the Fraunhofer minimum, where the transmission drops, the TWPA $\vert S_{11}\vert$ increases to similar levels as the maximum $\vert S_{21}\vert$ at low field.
Thus we can conclude that the TWPA mostly becomes reflective as expected from the model.

Looking closer at the reflection data, we see smaller moving ripples in the TWPA reflection moving as a function of $B_{\parallel,1}$.
They are also visible in the TWPA transmission, but reflection measurements are likely better for seeing small impedance mismatches.
It is conceivable that these features can be understood and thus that the disorder in the JJ array can be understood better from measuring a magnetic field dependence of the TWPA reflection because ultimately the Fraunhofer-like patterns contain information about the current profile inside the JJs. 
Thus, it is possible that magnetic field dependence of TWPAs can be an in-situ diagnostic that can help understand how to make more uniform JJ arrays.

We also looked for signs of the TWPA becoming more dissipative for the $B_\perp$ direction due to vortices in the JJ array or in the large bond pads at the input and output.
To this end, we compared the transmission and reflection data as a function of $B_\perp$ (not shown) with the data in \cref{fig:TWPA_transmission_vs_reflection}. 
However, we also see a similar level of $\vert S_{11}\vert$ for regions where $\vert S_{21}\vert$ is already suppressed, meaning the dominant effect with $B_\perp$ is also just reflection due to mismatched impedance.

\section{TWPA figures of merit and optimization procedure}
\label{sec:twpa_figs_of_merit}

\begin{figure}
  \centering
  \includegraphics[width=\columnwidth]{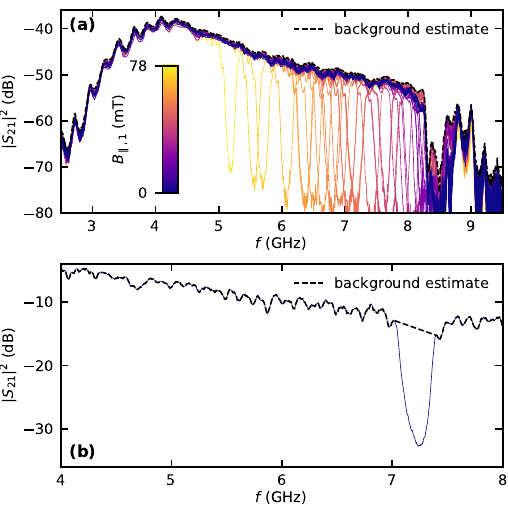}
%  \end{center}
  \caption{
      TWPA transmission spectrum for different $B_{\parallel,1}$ (colored lines).
      The data is taken from the main Fraunhofer lobe.
      The circulators are specified in the 4-\SI{8}{\giga\hertz} range, which is clearly visible in the $S_{21}$.
      We take the maximum $S_{21}$ at every frequency to estimate a background (black line) that we use for subtraction in the gain estimation.
      This background will more likely underestimate than overestimate the gain. 
    }
  \label{fig:background_estimation}
\end{figure}

\begin{figure}
  \centering
  \includegraphics[width=\columnwidth]{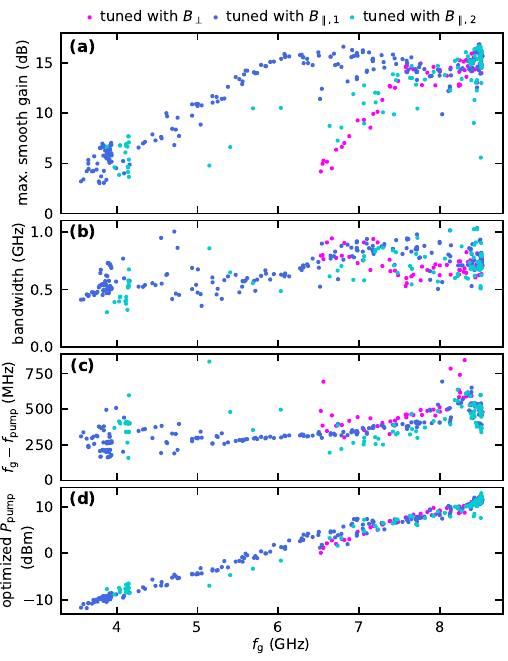}
%  \end{center}
  \caption{
    TWPA figures of merit as a function of the estimated $f_\mathrm{g}$ for $B_{\parallel, 1}$, $B_{\parallel, 2}$ and $B_\perp$.
    The $f_\mathrm{g}$ tuning is different for all field directions, but this figure shows that $f_\mathrm{g}$ alone does not determine performance.
    \textbf{(a)} shows the gain varies widely even for the same $f_\mathrm{g}$, while the bandwidth in \textbf{(b)} does not change dramatically.
    \textbf{(c)} and \textbf{(d)} show that the optimum $P_\mathrm{pump}$ and optimum $f_\mathrm{pump}$ are mainly a function of $f_\mathrm{g}$ and mostly independent of field direction.
    }
  \label{fig:figs_of_merit_different_Bs}
\end{figure}

A TWPA is an amplifier, therefore it has the usual figures of merit, e.g. gain, added noise (noise temperature or noise figure), bandwidth, and saturation power. 
For the scope of this work, we have focused on gain, bandwidth, gain fluctuations, and saturation power to understand how they change with magnetic field.
We took some measurements to estimate the saturation power, but mostly use $P_\mathrm{pump}$ as a proxy.
The setup in the fridge with the magnet was not prepared to accurately measure the added noise of the TWPA precisely, but the TWPA was characterized by Silent Waves before shipping (see \cref{tab:twpa_params} for a summary).

To estimate gain as a function of frequency, one would ideally do a full calibration of the setup and compare the $S_{21}$ with and without the optimized TWPA in the line. 
Our setup for the magnetic field dependence did not allow this and we did not take a calibration measurement without the TWPA that could be subtracted.
There are different ways to still estimate the gain, but they have limitations.
One can subtract the pump-off transmission from the pump-on transmission [cf. \cref{fig:fig1} \textbf{(d)}].
However, since the pump moves the TWPA bandgap (because it increases $L_\mathrm{J}$), this would lead to artifacts.
For TWPA A, we have instead taken the TWPA transmission spectra at different $B_{\parallel,1}$ within the first Fraunhofer lobe, which has the bandgap at different frequencies, and taken the maximum at every frequency to estimate a background without a bandgap [see \cref{fig:background_estimation} \textbf{(a)}].
Taking the maximum at every frequency makes it less likely to overestimate the gain, but reflection at the TWPA input is likely subtracted from the data as well.
This is the background we subtracted to get the gain curves [see for instance \cref{fig:fig3} \textbf{(a)}].
Crucially, we have always subtracted the same background, such that changes in the gain with magnetic field are relative to the same background. 
However, generally, the TWPA transmission does not change drastically with $B_{\parallel,1}$ [see \cref{fig:background_estimation} \textbf{(a)}] until it collapses in the Fraunhofer dip and eventually is suppressed above the critical field.
When the transmission is reduced as a function of field, we also find that the reflection at the TWPA increases [see \cref{sec:TWPA_transmission_vs_reflection}].
For TWPA B which was mounted inside the magnetic shields, we could not move the bandgap without compromising the TWPA, therefore we just linearly interpolated the background in the bandgap region [see \cref{fig:background_estimation} \textbf{(b)}] but also used the same background for all gain estimates.

With an established background to estimate the gain for each TWPA, we now describe the procedure to find the optimum $P_\mathrm{pump}$ and $f_\mathrm{pump}$ at different field settings.
The gain generally shows strong wiggles, likely due to impedance mismatches which could stem from deviations from the designed periodicity of the JJ array, reflections at the SMA connectors or the PCB, box modes from the package, and, in the magnetic field measurements, also our circulators being used at the edge of their specified range.
The position of the wiggles shifts with  $f_\mathrm{pump}$ and $P_\mathrm{pump}$, such that one could usually achieve $~\SI{20}{\decibel}$ of gain at a specific frequency within the larger bandwidth window but only at a more limited bandwidth of $~\SI{30}{\mega\hertz}$.
For usual use cases, such as qubit readout, one could also directly optimize on the signal-to-noise ratio at a concrete frequency.
However, for the field dependence we want to investigate, we decided that the optimum smooth gain would be a reasonable figure of merit to use as a cost function for optimization [see \cref{fig:fig3} \textbf{(a)}].
Alternatively, one could also define another figure of merit that penalizes gain fluctuations.
Given that we used the same optimization for the field dependence, we believe that the optimum smooth gain still offers a good way to compare the TWPA at different magnetic fields. 
The smoothed gain also allows for the definition of a relatively robust \SI{3}{\decibel} bandwidth.

We can now compare the three different field directions.
In \cref{fig:fig3} \textbf{(b)} and \cref{fig:fig4} \textbf{(d)} and \textbf{(e)}, the gain was plotted as a function of the respective field strength; by looking at the gain and bandwidth as a function of the $f_\mathrm{g}$, we show that even for the same $f_\mathrm{g}$, which tunes differently with $B_{\parallel,1}$, $B_{\parallel,2}$ and $B_{\perp}$, the TWPA performance differs strongly, see \cref{fig:figs_of_merit_different_Bs} \textbf{(a)} and \textbf{(b)}.
The only field direction for which the gain stays high over an extended field range is $B_{\parallel,1}$ as expected, because for $B_{\parallel,2}$ the bandgap closes and reopens, while for $B_\perp$ likely vortices lead to the strong dissipation.
The TWPA bandwidth is eventually reduced even for $B_{\parallel,1}$, the outliers in the bandwidth at high fields have much lower maximum gain which makes a larger bandwidth easier.

To understand the TWPA tune-up for a changing $f_\mathrm{g}$, we also consider the pump frequency $f_\mathrm{pump}$ and power $P_\mathrm{pump}$, obtained by optimizing the maximum smooth gain at every field, see \cref{fig:figs_of_merit_different_Bs} \textbf{(c)} and \textbf{(d)}.
We find that at low fields (high $f_\mathrm{g}$) the optimum $f_\mathrm{pump}$ is slightly further away from the center of the bandgap and it moves a bit closer as the fields tune down $f_\mathrm{g}$.
For the $P_\mathrm{pump}$ there seems to be a clear linear relationship between optimum $P_\mathrm{pump}$ and $f_\mathrm{g}$ independent of the field direction.
Intuitively, this is expected because the field reduces the critical current $I_\mathrm{c}$ of the JJs and hence the optimum $P_\mathrm{pump}$, as well as the saturation power of the TWPA (the saturation power of the TWPA is related to $P_\mathrm{pump}$ and therefore also likely linearly related to $f_\mathrm{g}$). 

\section{Setup and field dependence for the shielded TWPA}
\label{sec:TWPA_in_fridge_setup}

\begin{figure*}
  \centering
  \includegraphics[width=\textwidth]{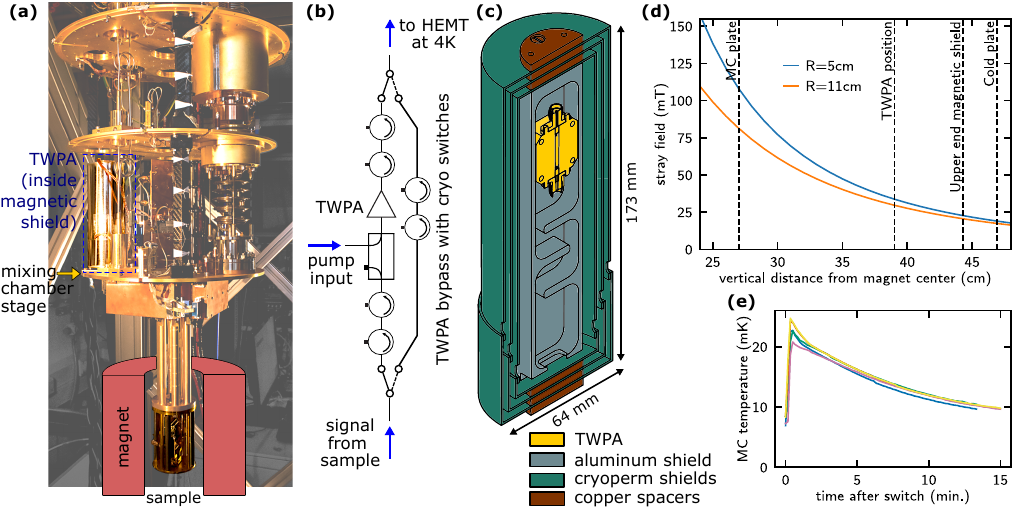}
%  \end{center}
  \caption{
    \textbf{(a)} Fridge photograph and wiring diagram of the TWPA.
    The TWPA inside the magnetic shields is mounted above the mixing chamber.
    It is sandwiched between two double-junction isolators (LNF-ISISC4\_12A) for ESD protection. 
    Two SPDT Switches (Radiall R571433141) were integrated into the wiring such that the TWPA can be bypassed.
    The bypass line has a separate set of circulators.
    \textbf{(b)} circuit diagram of the switch setup. 
    \textbf{(c)} colorized technical drawing of the magnetic shields and TWPA. 
    Three cylindrical cryoperm shields of 1mm strength are used.
    \textbf{(d)} stray field of the $B_\mathrm{z}$ magnet at \SI{6}{\tesla} as a function of vertical distance from the center of the magnet for different radial distances corresponding approximately to inner and outer shield boundaries.
    TWPA position and fridge stages are marked by dashed black lines.
    \textbf{(e)} MC temperature as a function of time after a switch event, measured several times to show consistency.}
  \label{fig:twpa_in_fridge}
\end{figure*}

In this section, we describe the mounting of the TWPA with magnetic shielding and cryogenic switches that allow bypassing the device. 
This setup then allows flexible use of the TWPA when needed up to the fields allowed by the shielding.
The setup is illustrated in \cref{fig:twpa_in_fridge}.
\cref{fig:twpa_in_fridge} \textbf{(a)} shows a photograph of the fridge with the TWPA mounted above the mixing chamber plate and the magnet and sample positions indicated.  
From the datasheet of the magnet, we can estimate the stray field at the TWPA position: at \SI{1}{\tesla}, the stray field would already be on the order of \SI{5}{\milli\tesla}, enough to severely compromise the TWPA. 
The stray field for \SI{6}{\tesla} at the sample as a function of distance is shown in \cref{fig:twpa_in_fridge} \textbf{(d)}.
Instead of using only passive magnetic fields, one can also use an additional compensation coil to cancel the stray field at the position of the TWPA~\cite{DiVora2023}.

With these stray fields in mind, we contacted a company to make a mu-metal-like shielding setup for the TWPA.
A schematic of the shields is shown in \cref{fig:twpa_in_fridge} \textbf{(c)}.
We chose 3 layers of \SI{1}{\milli\meter} thick Cryoperm - an iron-nickel alloy that undergoes special treatment. 
As an inner shield, we chose an aluminum box, which should act as a perfect diamagnet up to the critical field of bulk aluminum of around \SI{10}{\milli\tesla}.  
The shields have holes for the two coaxial cables for the TWPA input and output and for a copper braid that is used to thermally anchor the TWPA inside the shields directly to the mixing chamber.
Given that we have a bottom-loading fridge that is used for other experiments as well, it is crucial that the magnetic shields are not compromised by sweeping the field to \SI{6}{\tesla} because it would have meant opening the fridge to remove the shields and TWPA in case higher fields would be needed.

To preserve different functionalities in the fridge, we also installed current-operated cryogenic switches that allow bypassing the TWPA. 
This is useful, as the TWPA does not work at high fields, and because it only gives a signal-to-noise improvement in a frequency range that is narrow compared to the HEMT.
Also, when it is not pumped, it adds frequency-dependent attenuation to the line, as is evident from the wiggles in the pump-off transmission figures.
The wiring diagram is shown in \cref{fig:twpa_in_fridge} \textbf{(b)}.
There are isolators (model LNF-ISISC4\_12A) between the switches and the TWPA on both sides, that provide a low-impedance path to ground and protect the high-impedance TWPA from electrostatic discharge.
Both the bottom loading and as well as the switching could otherwise be dangerous to the TWPA.
By testing both, we confirmed that the TWPA is safe in this configuration.
We found a good current and time for reliable switching which does not raise the mixing chamber temperature much. 
Several example curves of the mixing chamber temperature as a function of time after switching events are shown in \cref{fig:twpa_in_fridge} \textbf{(e)}.
The temperature generally stays below \SI{25}{\milli\kelvin} and base temperature is reached again after about \SI{20}{\minute}.
Success of switching was confirmed each time.

\begin{figure}[bt]
  \centering
  \includegraphics[width=\columnwidth]{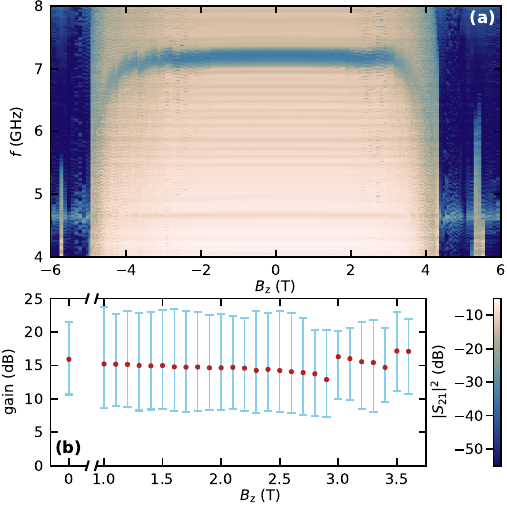}
%  \end{center}
  \caption{
    \textbf{(a)} TWPA $S_{21}$ as a function of field swept from \SI{6}{\tesla} to \SI{-6}{\tesla} for the shielded TWPA. 
    There is strong hysteresis in the data, likely originating from the magnetic shields. 
    \textbf{(b)} TWPA maximum smooth gain and gain range in the \SI{3}{\decibel}-bandwidth window at high fields. 
    Here, $f_\mathrm{pump}$ and $P_\mathrm{pump}$ were optimized at higher fields. 
    }
  \label{fig:wide_range_shielded_field_sweep}
\end{figure}

In the main text, $\vert S_{21}\vert$ data for the shielded TWPA is only shown up to \SI{3.5}{\tesla} and $f_\mathrm{gap}$ starts to drop significantly for $B_\mathrm{z}>\SI{2.5}{\tesla}$ (see \cref{fig:fig5}).
We did not sweep the field to \SI{6}{\tesla} right away, because we were initially worried that at the highest fields, the magnetic shields might be magnetized which would not be reversible while they remain inside the dilution refrigerator.
However, ultimately we swept the magnetic field over the maximum range. 
The TWPA $\vert S_{21}\vert$ as a function of $B_\mathrm{z}$ for the downsweep from \SI{6}{\tesla} to \SI{-6}{\tesla} can be found in \cref{fig:wide_range_shielded_field_sweep} \textbf{(a)}.
It is evident, that the TWPA transmission breaks down relatively suddenly for the highest fields and that there is hysteresis on the order of \SI{0.5}{\tesla}.
The $f_\mathrm{g}$ exhibits a plateau between approximately \SI{2}{\tesla} and \SI{-2}{\tesla}.
From the stray field data [\cref{fig:twpa_in_fridge} \textbf{(d)}], we can estimate the stray field at the TWPA position for \SI{2}{\tesla} at the magnet center to be approximately \SI{11}{\milli\tesla} if the shields were not present.
Outside it shows the expected reduction with field, but also some jumps in particular between \SI{-2}{\tesla} and \SI{-4}{\tesla}.
We possibly see a Fraunhofer lobe both at positive and negative fields. 
The field direction is likely a superposition of $B_{\parallel,2}$ and $B_\perp$ in the TWPA coordinate system.
While the hysteresis and sudden breakdown of the $\vert S_{21}\vert$ are reminiscent of the $B_\perp$ dataset of the unshielded TWPA, both the sudden breakdown and hysteresis are likely due to the magnetic shields.

For the gain measurement as a function of $B_\mathrm{z}$ for the shielded TWPA, shown in \cref{fig:fig5} \textbf{(b)}, we did not adapt $f_\mathrm{pump}$ and $P_\mathrm{pump}$, because we were interested in the performance of the TWPA in the most straightforward mode of operation.
\cref{fig:wide_range_shielded_field_sweep} \textbf{(b)} shows additional data, where it was attempted to optimize $f_\mathrm{pump}$ and $P_\mathrm{pump}$ at higher fields.
A good $f_\mathrm{pump}$ guess and a corresponding $P_\mathrm{pump}$ can usually be based on $f_\mathrm{g}$, but above \SI{3}{\tesla} the bandgap becomes less visible and slightly above \SI{3.5}{\tesla} no good gain could be realized.
However, in the unpumped $\vert S_{21}\vert$, there is already more noise visible above \SI{2}{\tesla} and this also shows up in the gain curves, leading to visibly more added noise. 
We would therefore conclude that the valid operating window for the TWPA in our setup is roughly \SI{-2}{\tesla} to \SI{2}{\tesla}.

\section{TWPA temperature dependence}
\label{sec:TWPA_vs_T}

\begin{figure}[tbh]
  \centering
  \includegraphics[width=\columnwidth]{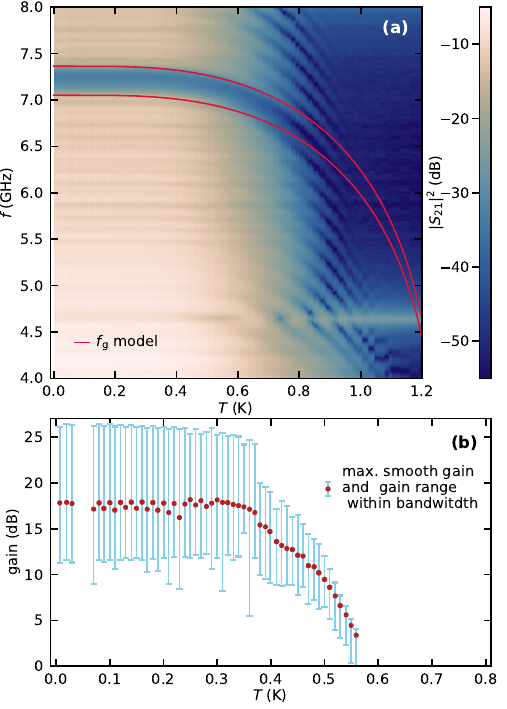}
%  \end{center}
  \caption{
    \textbf{(a)} Temperature dependence of the TWPA transmission. The bangap is modeled using the TWPA B parameters in \cref{tab:twpa_params}, and a temperature-dependent gap based on \cref{eq:Delta_vs_T} with  $T_\mathrm{c}=\SI{1.27}{\kelvin}$.  
    \textbf{(c)} Best values for optimized maximum smooth gain at different temperatures. The error bars show the maximum and minimum gain inside the \SI{3}{\decibel} bandwidth. 
    Up to \SI{0.3}{\kelvin} there is no strong temperature dependence, neither for the transmission nor for the maximum gain.
    }
  \label{fig:TWPA_vs_T}
\end{figure}

We investigate here the temperature dependence of the TWPA response.
If the TWPA could be mounted at a higher, warmer temperature stage of the fridge, that would increase the distance to the magnet and decrease the stray field [see \cref{fig:twpa_in_fridge} \textbf{(d)}].
Data of the TWPA transmission and optimized gain as a function of temperature are shown in \cref{fig:TWPA_vs_T}.
Different temperatures were stabilized using a PID controller on the mixing chamber heater with the mixing chamber temperature as the signal.
We stabilized each temperature for about \SI{10}{\minute} to ensure thermalization of the TWPA and rejected datasets where temperature fluctuations were too high. 

Apart from mounting the TWPA away from a magnet, the temperature dependence is interesting to compare to the magnetic field dependence, because both suppress the superconducting gap. 
For a BCS superconductor such as aluminum, a good approximation (within less than 2~\%) for the temperature-dependent gap is~\cite{Gross1986}:
\begin{equation}
\Delta(T) = \Delta(0) \tanh \left ( 1.74 \sqrt{T_\mathrm{c}/T-1} \right ),
\label{eq:Delta_vs_T}
\end{equation}
where we estimate the critical temperature $T_\mathrm{c}=\SI{1.27}{\kelvin}$ based on the thicker  \SI{50}{\nano\meter} aluminum film which would have the smaller $T_\mathrm{c}$~\cite{Marchegiani22}.
Using this gap dependence and the TWPA B parameters from \cref{tab:twpa_params}, we can model the temperature dependence of $f_\mathrm{g}$ analogous to the field dependence described in \cref{sec:twpa_model}. 
Data and model are shown in \cref{fig:TWPA_vs_T} \textbf{(a)}.
The model describes the evolution of $f_\mathrm{g}$ well up to about \SI{0.6}{\kelvin}. 
For higher temperatures, the bandgap fades as overall transmission is suppressed.
Instead of $\vert S_{21} \vert$ showing a clear boundary as $f_\mathrm{p}$ is reduced with $T$, which is the case for the magnetic-field dependence,  $\vert S_{21} \vert$ instead develops ripples that move down with temperature. 

Up to \SI{0.3}{\kelvin}, the almost unchanged $\vert S_{21}\vert$ and gain suggests that the TWPA is not strongly compromised, but we did not explicitly measure the added noise at each temperature.
Given that the TWPA added noise is likely on the order of \SI{0.4}{\kelvin}~\cite{Planat2020}, it is not implausible that raising the device temperature at first does not increase the added noise too much. 
However, we do not claim that the TWPA has the same performance at \SI{0.3}{\kelvin} as at base temperature, just that our measurement of gain and device transmission do not show a difference.
We conclude that one could likely mount the TWPA at the cold plate of the dilution refrigerator which is at \SI{0.1}{\kelvin}.
This could increase the distance to the magnet and likely at least double the maximum field at which the TWPA can be operated without adding more shielding.

% \bibliography{References_cQED}

\end{document}